\documentclass[a4paper,fleqn,usenatbib]{mnras}

\usepackage{newtxtext,newtxmath}

\usepackage[T1]{fontenc}
\usepackage{ae,aecompl}

\usepackage{graphicx}	
\usepackage{amsmath}	
\usepackage{amssymb}	

\newcommand{\se}[1]{\S\ref{sec:#1}}

\newcommand{\Fig}[1]{Figure~\ref{fig:#1}}
\newcommand{\Figs}[1]{Figures~\ref{fig:#1}}

\newcommand{\be}{\begin{equation}}
\newcommand{\ee}{\end{equation}}
\newcommand{\bea}{\begin{eqnarray}}
\newcommand{\eea}{\end{eqnarray}}

\newcommand{\Msun}{M_\odot}

\newcommand{\ifm}[1]{\relax\ifmmode#1\else$\mathsurround=0pt #1$\fi}
\newcommand{\kms}{\ifmmode\,{\rm km}\,{\rm s}^{-1}\else km$\,$s$^{-1}$\fi}
\newcommand{\hmpc}{\,\ifm{h^{-1}}{\rm Mpc}}

\newcommand{\Mpc}{\,{\rm Mpc}}

\newcommand{\Gyr}{\,{\rm Gyr}}

\newcommand{\Myr}{\,{\rm Myr}}
\newcommand{\GyrI}{\,{\rm Gyr}^{-1}}
\newcommand{\sSFR}{\,{\rm sSFR}}

\newcommand{\ltsima}{$\; \buildrel < \over \sim \;$}
\newcommand{\lsim}{\lower.5ex\hbox{\ltsima}}
\newcommand{\gtsima}{$\; \buildrel > \over \sim \;$}
\newcommand{\gsim}{\lower.5ex\hbox{\gtsima}}

\def\la{\langle}

\def\omm{\Omega_{\rm m}}

\def\omb{\Omega_{\rm b}}

\def\tu{\,{\rm t}_{\rm U}}

\def\Mv{M_{\rm vir}}

\def\Ms{M_*}

\usepackage{color}

\title[FL II]{FirstLight II: Star formation rates of primeval galaxies from z=5-15}

\author[Ceverino et al.]{
Daniel Ceverino,$^{1}$\thanks{E-mail: ceverino@uni-heidelberg.de}
Ralf S. Klessen,$^{1,2}$ Simon C. O. Glover,$^{1}$
\\
$^{1}$Universität Heidelberg, Zentrum für Astronomie, Institut für Theoretische Astrophysik, Albert-Ueberle-Str. 2, 69120 Heidelberg, Germany\\
$^{2}$Universität Heidelberg, Interdisziplinäres Zentrum für Wissenschaftliches Rechnen, INF 205, 69120, Heidelberg, Germany}

\date{Accepted XXX. Received YYY; in original form ZZZ}

\pubyear{2018}

\begin{document}
\label{firstpage}
\pagerange{\pageref{firstpage}--\pageref{lastpage}}
\maketitle

\begin{abstract}

In the FirstLight project, we have used $\sim$300 cosmological, zoom-in simulations to determine the star-formation histories of distinct first galaxies with stellar masses between $\Ms=10^6$ and $3 \times 10^9 \Msun$ during cosmic dawn ($z=5-15$). 
The evolution of the star formation rate (SFR) in each galaxy is complex and diverse, characterized by bursts of star formation.
Overall, first galaxies spend 70\% of their time in SF bursts.
A sample of 1000 of these bursts indicates that the typical burst at $z\simeq6$ has a specific SFR (sSFR) maximum of  $5-15 \GyrI$ with an  effective width of $\sim$100 Myr, one tenth of the age of the Universe at that redshift.
A quarter of the bursts populate a tail with very high  sSFR maxima of $20-30 \GyrI$ and significantly shorter timescales of $\sim$40-80 Myr.
This diversity of bursts sets the mean and the mass-dependent scatter of the star-forming main sequence.
This scatter is driven by a population of low-mass, $\Ms \leq 10^8 \Msun$, quiescent galaxies.
The mean  sSFR and the burst maximum at fixed mass increase with redshift, with the later always being a factor $\sim$2 higher than the former.
This implies  sSFR maxima of $\sim 20 - 60 \GyrI$ at $z=9-10$.
The SFR histories are publicly available at the FirstLight website.

\end{abstract}

\begin{keywords}
galaxies: evolution -- galaxies: formation  -- galaxies: high-redshift 
\end{keywords}


\section{Introduction}

Current surveys from the Hubble Space Telescope (HST) have yielded a population of primeval galaxies at cosmic dawn and reionization epochs, redshifts $z=5-10$ \citep{Bouwens04, Finkelstein12, Oesch13, Bouwens15}. 
Little is known about the properties of these first galaxies.
One of the most important properties is the rate at which gas is converted into stars, the star formation rate (SFR). 
This rate drives  galactic growth and galaxy evolution, as well as the reionization of the entire Universe.

The relation between the SFRs and stellar masses of galaxies at these epochs \citep{Duncan14, Salmon15} gives some clues about the physical mechanisms that shape the star formation (SF) histories at high redshifts.
A tight relation as seen at lower redshifts, $z\leq3$  \citep{Daddi07, Pannella09, Magdis10, Sawicki12, Steinhardt14} implies smooth SF histories, driven by the smooth accretion of gas into galaxies \citep{Keres05,  Dekel09, Dekel13, Bouche10, Goerdt15}.
In contrast, discrete bursts of SF set the scatter around the SF main sequence \citep{Shapley05, Mannucci09, Wyithe14}.
Therefore, the properties of this main sequence are set by the smooth and bursty modes of SF.

Analytical models, also called 'bathtub' models of galaxy formation \citep{Bouche10, Dave12, Lilly13} usually link the specific star formation rate, sSFR=SFR/$\Ms$, with the specific inflow rate of gas into dark-matter halos, modulated by the feedback-driven outflow rate \citep{NeisteinDekel08, Dekel13, DekelMandelker14}.
Using the Extended Press-Schechter (EPS) theory \citep{Bond91}, these models predict $\sSFR \propto (1+z)^{5/2}$ 
plus a mild mass dependence.
The 5/2 power can be understood from the following scaling argument of the EPS approximation.
In this theory, there is a self-invariant time variable, $\omega \propto D(a)^{-1}$, where $D(a)$ is the growth rate of linear density perturbations and $a=(1+z)^{-1}$. This time invariant implies that the halo mass growth, dM/d$\omega$, is constant and therefore $\dot{M} \propto \dot{\omega}$. In the Einstein-deSitter regime, valid at high redshifts, $a \propto t^{2/3}$, $D(a) \propto a$, and therefore,  
 $\sSFR \propto  \dot{M} / M \propto a^{-5/2}$. 
Cosmological simulations \citep{Dave11} and semi-analytical models \citep{SomervilleDave15} confirm this analytical prediction,
although their temporal resolution is low, which may lead to overly smooth sSFR histories.
On the other hand, early estimates of the sSFR at high redshift show little evolution \citep{Gonzalez11}.
However, they overestimate the stellar mass due to photometric contamination by emission lines \citep{Stark13}. 
Current observations agree with the theoretical predictions \citep{Salmon15},
although the old estimates are still used \citep{Behroozi13, Moster17}.

The evolution of the mean sSFR with redshift  is set by the combination of individual sSFR histories of many galaxies. Each history does not necessarily need to follow the mean trend. 
This generates an intrinsic scatter driven by variations in the individual SF histories, such as SF bursts \citep{Tacchella13, Mason15}.
Current simulations \citep{Dave13} and semi-analytical models underpredict the scatter of the relation \citep{Somerville08, Lu14} at $z=4-6$ by a factor 2-3 \citep{Salmon15}. 
This is mostly due to temporal and spatial resolution effects that artificially smooth sudden increases of the SFR driven by galaxy mergers or clumpy gas accretion. 


In this paper, we aim to characterize the sSFR histories, their SF bursts and their evolution during cosmic dawn.
Properties like the sSFR height of the burst and their typical duration  give some clues about the physical mechanisms responsible for these high redshift starbursts.
In addition, the frequency of these bursts for different galaxy masses and redshifts set their importance for the overall galaxy growth at these early epochs.


In order to characterize the mean properties and diversity of the sSFR histories, a large sample of high-resolution simulations, such as the FirstLight simulations, is needed.
The FirstLight database of cosmological zoom-in simulations of first galaxies reproduces the galaxy scaling relations, the UV luminosity function, and the galaxy stellar mass function in agreement with current observations \citep[][Paper I]{CGK}.
This database also predicts a rapid evolution of the power-law slope of the UV luminosity function, which reaches $\alpha\simeq-2.5$ at $z=10$, consistent with current estimates \citep{Oesch17}.
These predictions will be confirmed in future deep surveys by the James Webb Space Telescope (JWST) and large ground-based 30 m telescopes coming in the next decade.
Meanwhile, the publicly available FirstLight\footnote{\url{http://www.ita.uni-heidelberg.de/~ceverino/FirstLight}} database can be used to design successful JWST proposals.


The outline of this paper is as follows. Section \se{IC} describes the first data release of FirstLight.
Section \se{sSFR} gives some examples of sSFR histories.
Section \se{SFR} describes the star formation main sequence at $z\simeq6$.
Section \se{Bursts} is devoted to the properties of the SF bursts and their evolution is discussed in Section \se{evo}.
Section \se{summary} finishes with the conclusion and final discussions.

\section{The Simulations}
\label{sec:IC}

\begin{figure}
	\includegraphics[width=\columnwidth]{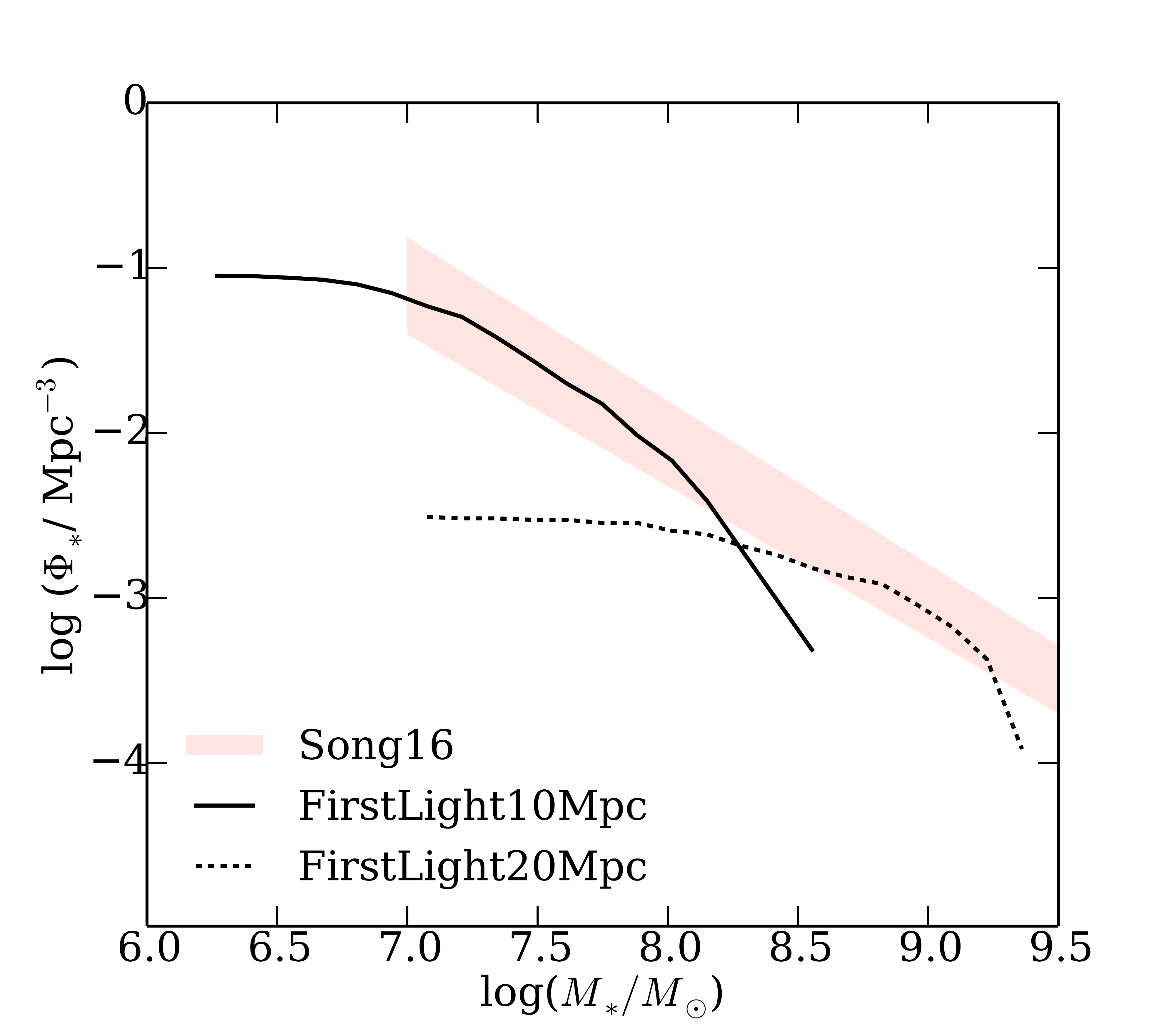}
	 \caption{Stellar mass functions of the FirstLight galaxies from the two cosmological volumes at $z=6$.  The combination of both samples agrees well with current observations \citep{Song16}.   }
	  \label{fig:SMF}
\end{figure}

\begin{figure*}
	\includegraphics[width=0.99 \textwidth]{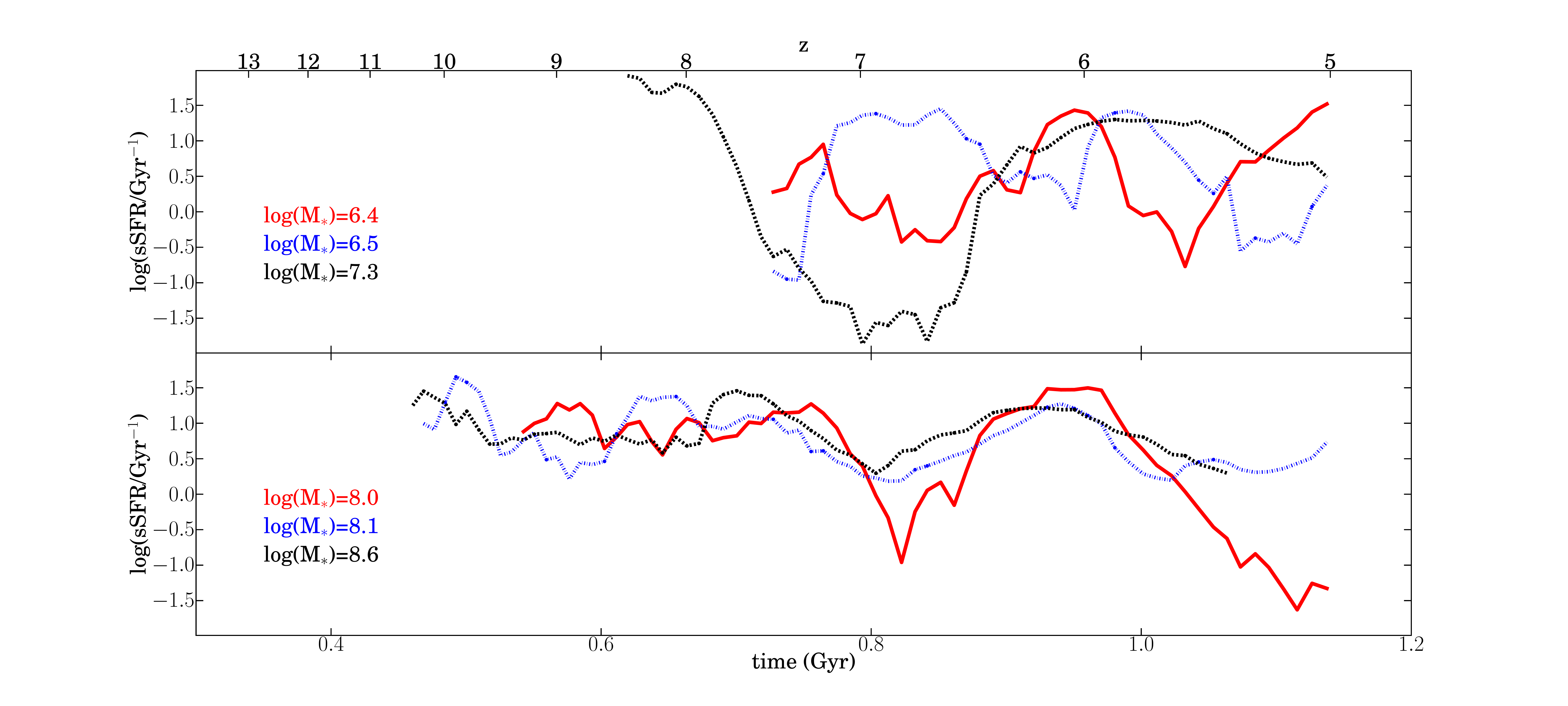}	
	  \caption{Evolution of the sSFR of selected galaxies with a burst of star formation that reaches $\sSFR_{\rm max}>15  \GyrI$  at $z\simeq6$ (t$_{\rm U}\simeq 1$ Gyr).}
	  \label{fig:SSFHStrong}
\end{figure*}
	  
\begin{figure*}
	\includegraphics[width=0.99 \textwidth]{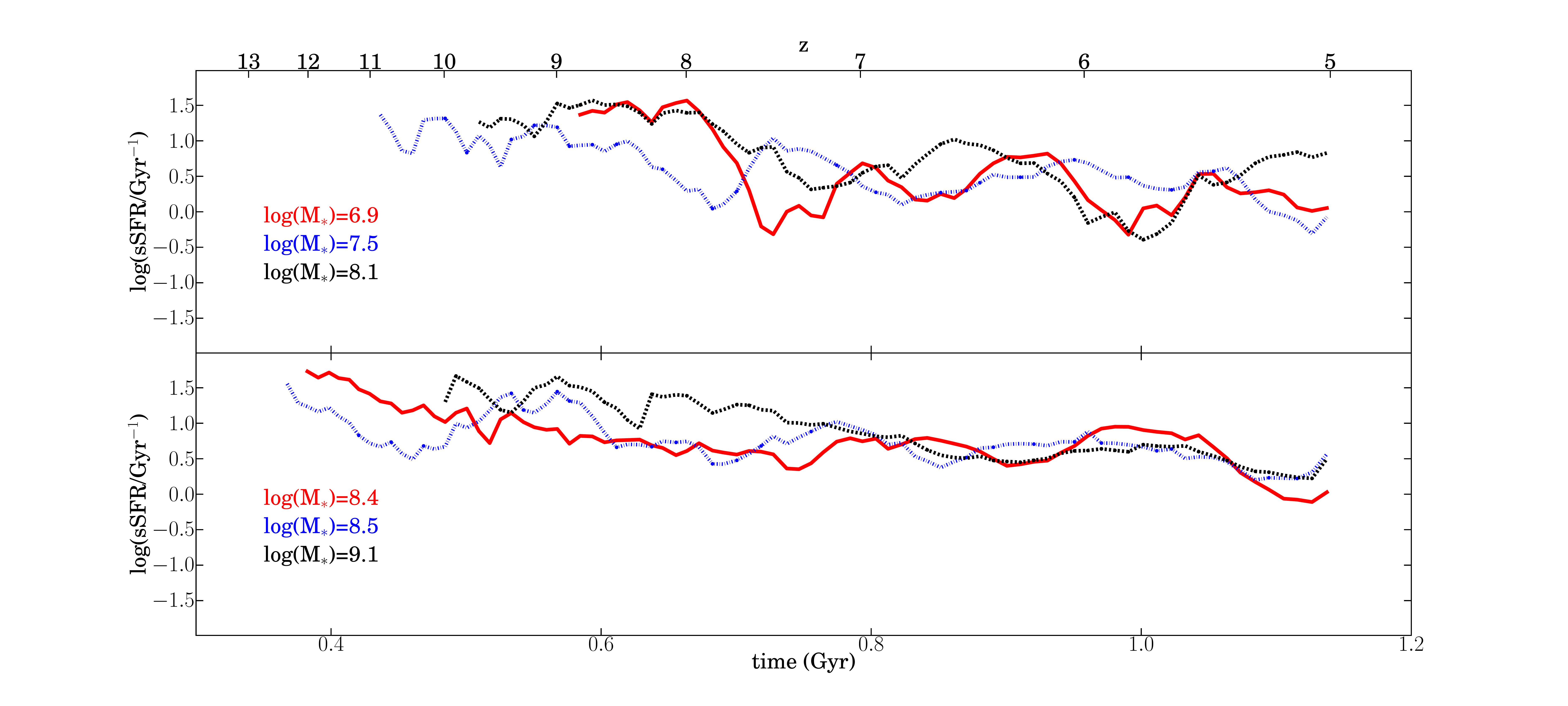}
	  \caption{Evolution of the sSFR of selected galaxies with a burst of star formation that reaches $\sSFR_{\rm max}=5-10  \GyrI$  at $z\simeq6$.}
	  \label{fig:SSFHWeak}
\end{figure*}

This paper uses a mass-selected subsample of
galaxies simulated in the FirstLight project described fully in paper I.
The subsample consists of 290 halos with a maximum circular velocity, V$_{\rm max}$, between 50 and 250 $\kms$, selected at $z=5$. The halos cover a mass range between a few times $10^9$ and a few times $10^{11} \ \Msun$. 
This range excludes more massive and rare halos with number densities lower than $\sim 3 \times 10^{-4} (h^{-1} \Mpc)^{-3}$, 
as well as small halos in which galaxy formation is extremely inefficient.
 
The target halos are initially selected using low-resolution N-body only simulations of two cosmological boxes with sizes 10 and 20 $\hmpc$, assuming WMAP5 cosmology with $\omm=0.27$, $\omb=0.045$, $h=0.7$, $\sigma_8=0.82$ \citep{Komatsu09}.  
We select all distinct halos with a maximum circular velocity (V$_{\rm max}$) at z = 5 greater than a specified threshold, log $V_{\rm cut}=1.7$ in the 10 $\hmpc$ box and log $V_{\rm cut}=2.0$ in 20 $\hmpc$ box.
Initial conditions for the selected halos with much higher resolution are then generated using a standard zoom-in technique  \citep{Klypin11}.
 The DM particle mass resolution is m$_{\rm DM}=10^4 \ \Msun$. The minimum mass of star particles is $100 \ \Msun$.
 The maximum spatial resolution is always between 8.7 and 17 proper pc (a comoving resolution of 109 pc after $z=11$).
 
The V$_{\rm max}$-selected sample covers more than 3 orders of magnitude in stellar mass, $\Ms=10^6-10^{9.5} \ \Msun$, although it probably misses some galaxies at both ends due to the intrinsic scatter in the V$_{\rm max}-\Ms$ relation.
80\% of the selected halos reach $z=6$. The corresponding galaxy stellar mass functions from the two cosmological volumes are shown in \Fig{SMF}. The combination of both samples agrees well with current observations \citep{Song16}.  
  
 The simulations are performed with the  \textsc{ART} code
\citep{Kravtsov97,Kravtsov03}, which accurately follows the evolution of a
gravitating $N$-body system and the Eulerian gas dynamics using an adaptive mesh refinement (AMR) approach.
Besides gravity and hydrodynamics, the code incorporates 
many of the astrophysical processes relevant for galaxy formation.  
These processes, representing subgrid 
physics, include gas cooling due to atomic hydrogen and helium, metal and molecular 
hydrogen cooling, photoionization heating by a constant cosmological UV background with partial 
self-shielding, star formation and feedback (thermal+kinetic+radiative), as described in paper I.

In short, star formation is assumed to occur at densities above a threshold of 1 cm$^{-3}$ and at temperatures below $10^4$ K. The code implements a stochastic star formation model that yields the empirical Kennicutt-Schmidt law \citep{Schmidt, Kennicutt98}.
In addition to thermal energy feedback, the simulations use radiative feedback, as a local approximation of radiation pressure. This model adds non-thermal pressure to the total gas pressure in regions where ionizing photons from massive stars are produced and trapped. The model of radiative feedback used is named RadPre\_IR in Ceverino et al. (2014) and it uses a a moderate trapping of infrared photons. 
The latest feedback model also includes the injection of momentum coming from the (unresolved) expansion of gaseous shells from supernovae and stellar winds \citep{OstrikerShetty11}.  More details can be found in Paper I, \citet{Ceverino09}, \citet{CDB}, and \citet{Ceverino14}. 

\section{Specific Star-Formation Rates of Selected Galaxies}
\label{sec:sSFR}

\begin{figure}
	\includegraphics[width=\columnwidth]{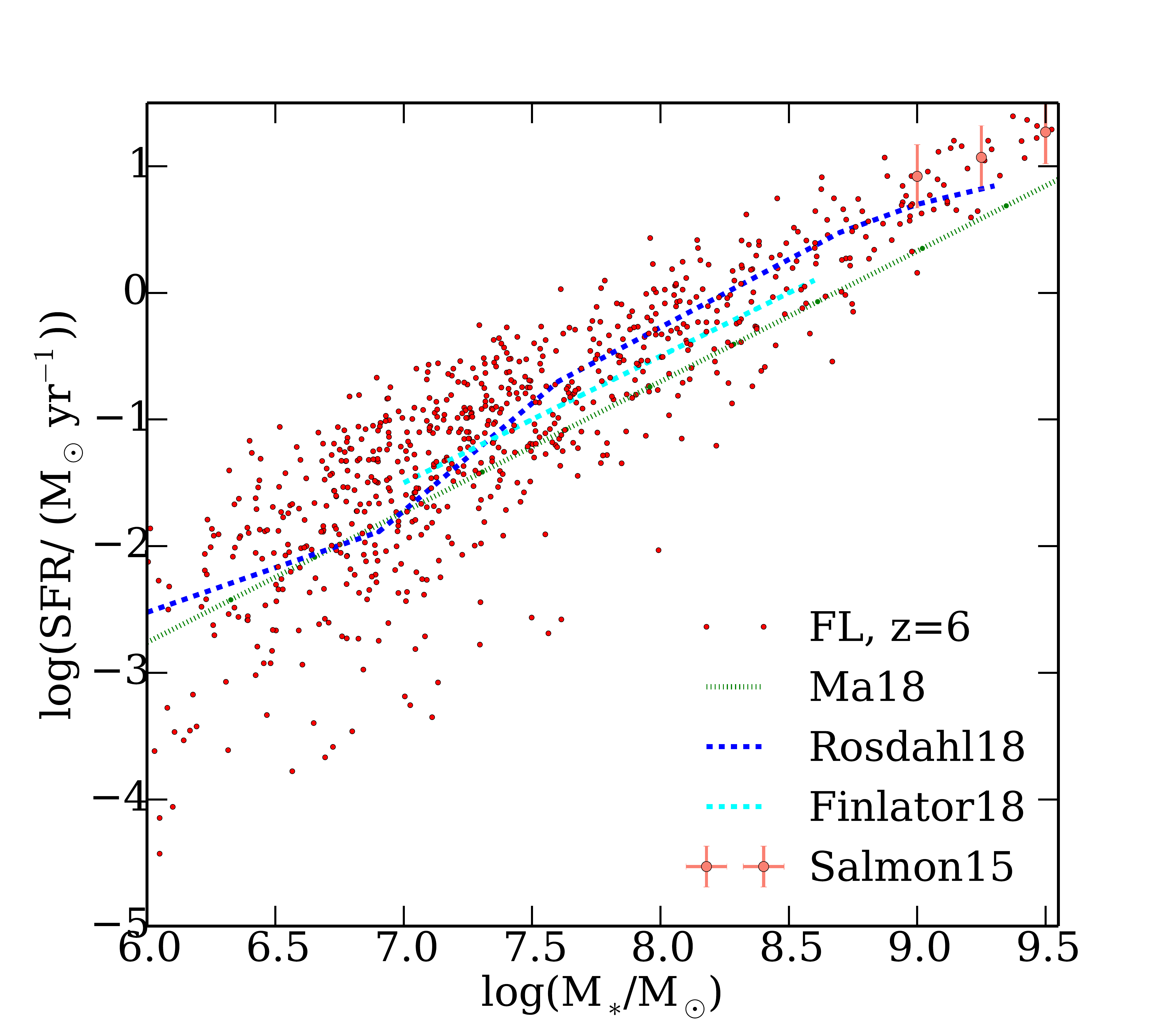}
	\includegraphics[width=\columnwidth]{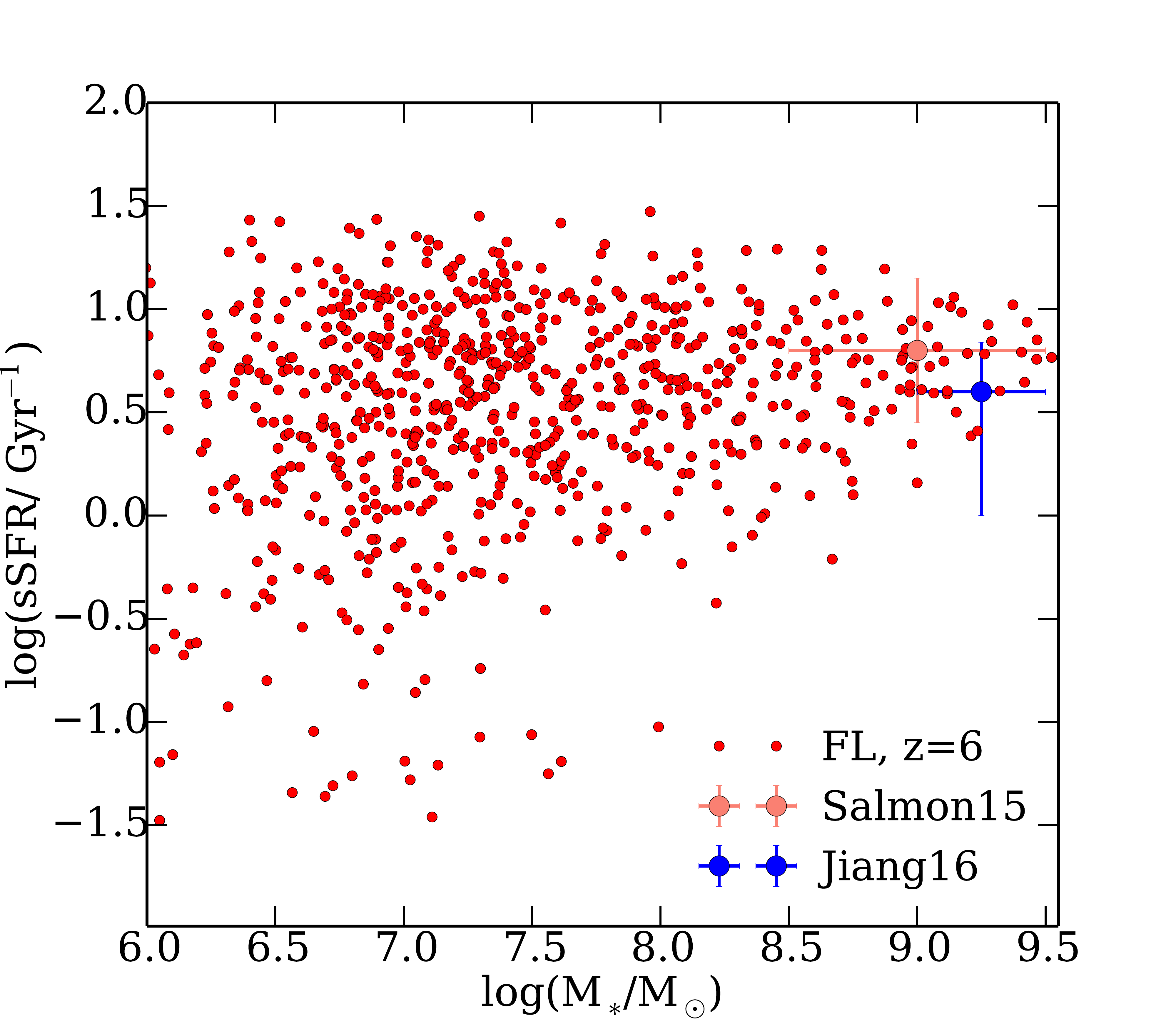}	
	 \caption{Top: SFR versus stellar mass ($\Ms$). The high-mass end is consistent with current observations. Bottom: sSFR versus $\Ms$. A mass-dependent scatter is driven by a population of low-mass, $\Ms \leq 10^8 \Msun$, quiescent galaxies.}
	  \label{fig:MS}
\end{figure}

\begin{figure}
	\includegraphics[width=\columnwidth]{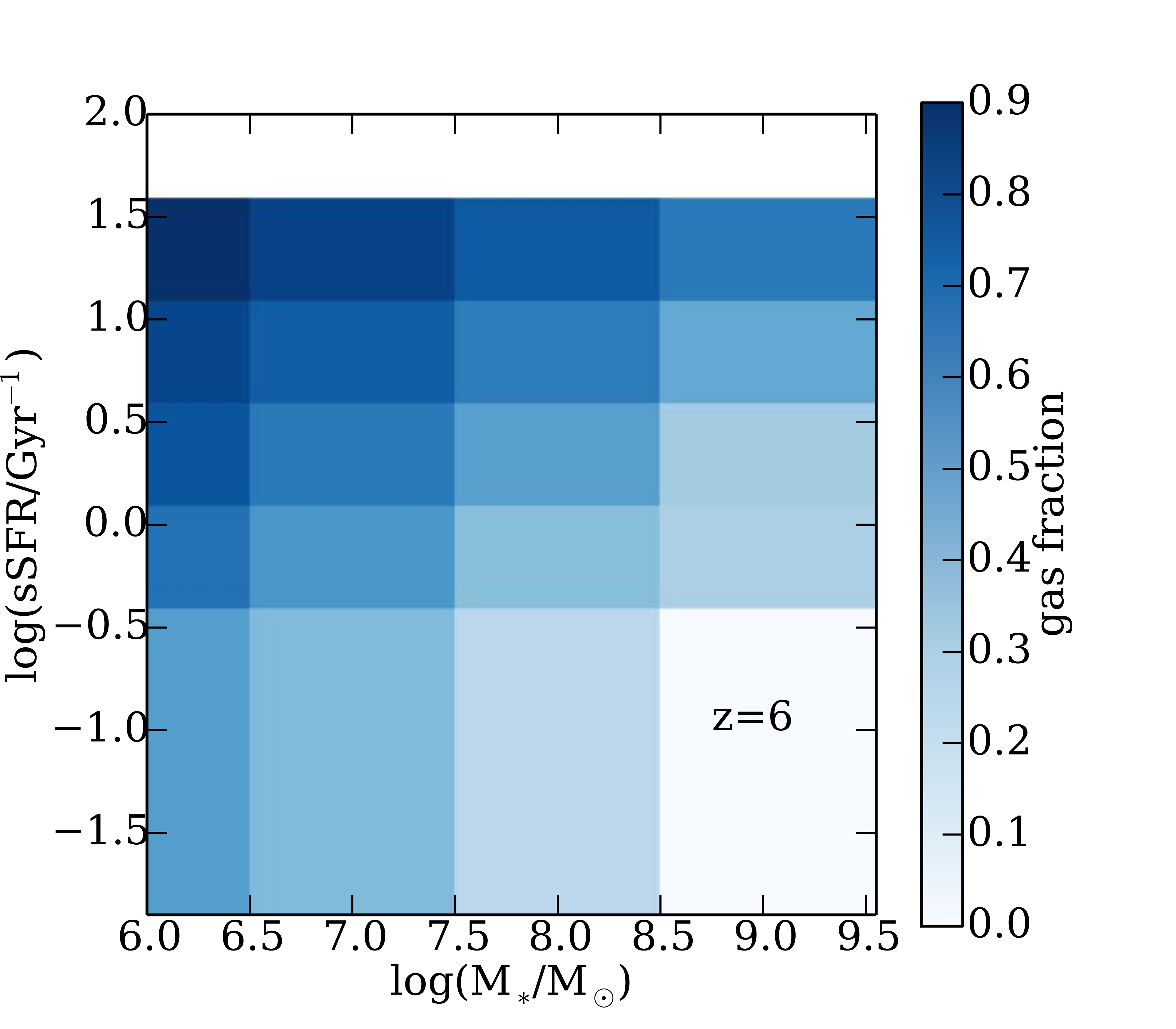}
	\includegraphics[width=\columnwidth]{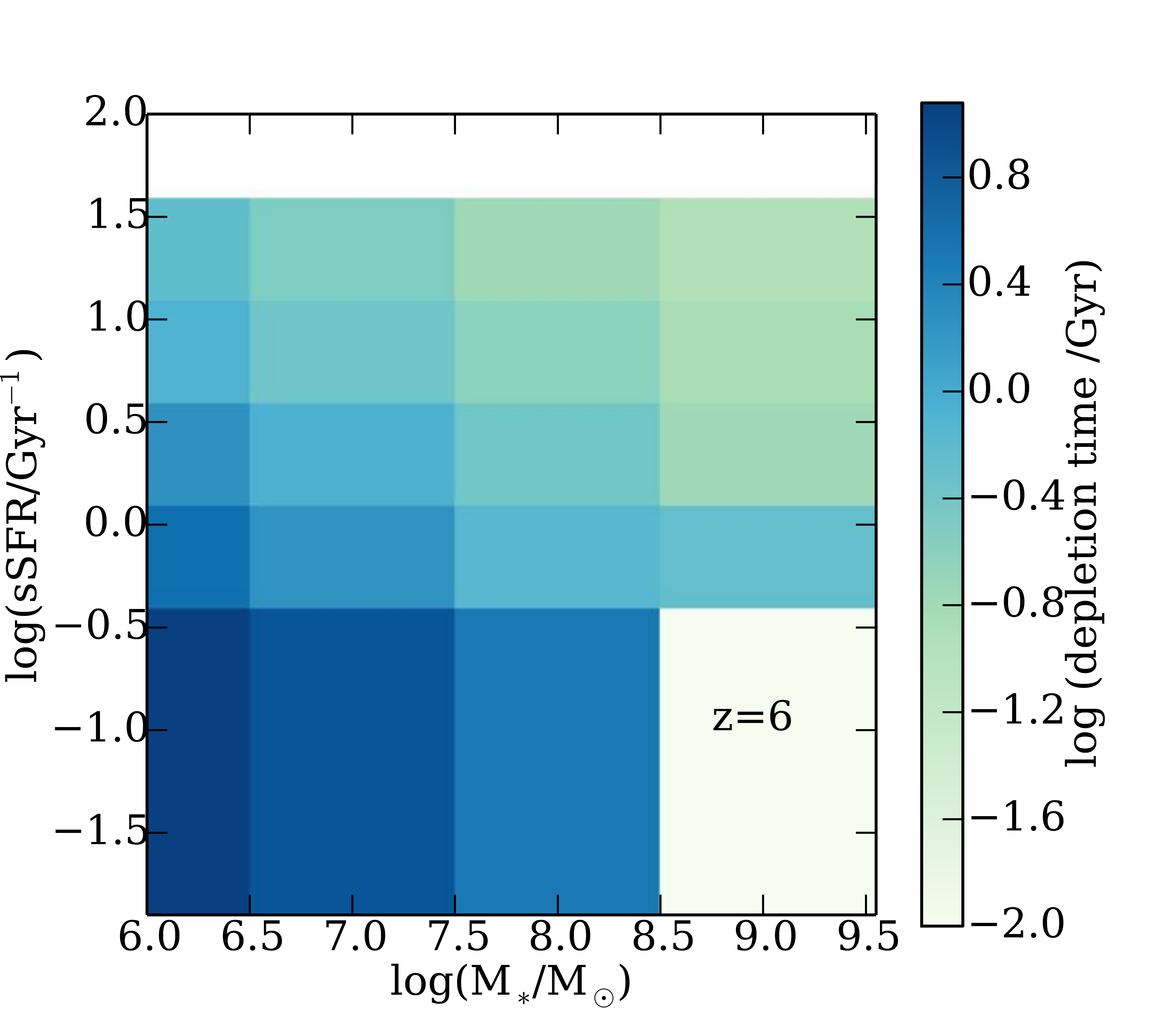}

	 \caption{Top: gas fraction. The gas fraction decreases with increasing galaxy mass and increases with sSFR.
A high gas content is therefore needed to generate high sSFR at fixed mass. 
Bottom: gas depletion time.  The depletion time is shorter at higher sSFRs, 
driven by violent and dissipative processes}
	  \label{fig:tD}
\end{figure}

Representative examples of the diversity of sSFR histories are shown in \Figs{SSFHStrong} and \ref{fig:SSFHWeak}. They cover the range of galaxy masses of the FirstLight database, from $\Ms\simeq 10^{6.5} $ to $10^{9.5} \Msun$ at $z=6$, when the Universe is roughly $\tu=1 \Gyr$ old.
Each of the $\sim300$ sSFR histories starts when the major (most massive) progenitor reaches a virial mass of $\Mv=10^9 \ \Msun$.
It ends in the last available snapshot at $z\ge5$. Overall, each history extends over $300-800$ Myr, depending on the particular galaxy.
The SFR of the main progenitor at each snapshot is computed using the star particles younger than $\sim 10 \Myr$.

Individual galaxies show a complex evolution of their sSFR with time. This is very different from the smooth evolution predicted by EPS theory \citep{NeisteinDekel08, DekelMandelker14}. 
Typical sSFR histories are characterized by star-formation bursts,
defined as a peak in the sSFR history around each local maximum. These bursts 
  can last a few hundreds Myrs, a significant fraction of the age of the Universe at these high redshifts. During these bursts, the sSFR may vary upto more than 2 orders of magnitude. 
Overall, galaxies spend $\sim$70\% of the time in these bursts, independent of their final mass.
 \Fig{SSFHStrong} shows different examples of SF bursts that can  reach $\sSFR\simeq20 \GyrI$ at $z\simeq6$.   After the peak, the  sSFR typically declines to very low values, $\sSFR < 0.3/\tu$, below the threshold for quiescent galaxies at lower redshifts  \citep{Damen09}. 
Galaxies do not remain in this quiescent phase for long, because frequent peaks of SF generate a stochastic, burst-like history. 

We find and characterize the SF bursts in the following way.
First, we smooth each sSFR history with a median filter of 5 points.  This limits our analysis to SF bursts longer than $\sim30$ Myr. As shown below, the typical duration of a burst is indeed much longer than this. 
We use a continuous wavelet analysis to find all sSFR maxima.
Overall, we find $\sim$1000 maxima or bursts between redshifts $z=5-15$.
The total duration of each SF burst is defined as the period of time between the two minima at both sides of each local maximum.
The full-width-half-maximum (FWHM) around the SF peak is another indicator of the effective width of the burst.
Finally, we look at the time between two consecutive sSFR maxima.

A careful analysis of  the sSFR histories shows that not all galaxies experience the SF bursts shown in \Fig{SSFHStrong}. In fact, many galaxies have smoother histories with SF peaks in the range of $\sSFR=5-15 \GyrI$ at $z\simeq6$ (\Fig{SSFHWeak}).
These typical bursts are followed by a decrease in the sSFR but the sSFR minima is not as low as in the previous examples ($\sSFR \ge 0.3/\tu$).
In summary, first galaxies show a large diversity of SF histories. Only a large sample of high-resolution simulated galaxies is able to characterize this diversity and the scatter in the SF main sequence at $z \ge 6$.

\section{The Star-Forming Main-Sequence at redshift 6}
\label{sec:SFR}

A clear star-forming main sequence (SFMS) relation  is present in the sample at $z\simeq6$ (\Fig{MS}). 
We use three different snapshots at $z=5.5$, 6 and 6.5 in order to mimic the typical redshift binning used in observations. Due to the variations in the SFR of individual galaxies, we can consider each snapshot as a different galaxy. 
This oversampling does not bias our results.
The high-mass end ($\Ms\simeq10^9 \ \Msun$) is consistent with current observations.
Its mean and scatter agree remarkably well with observations \citep{Salmon15, Jiang16}.
The scatter around the SFMS mean is $\sigma=0.3$ dex. 
However, we note that it is
much higher than the values from other cosmological simulations and semi-analytical models  \citep{Finlator11, Dave13, Lu14, Somerville08} that show a scatter of around $\sigma\simeq0.1$ dex. 
This disagreement is most probably due to the much lower temporal and spatial resolution of these earlier models.
The FirstLight simulations are able to resolve intense SF bursts on timescales smaller than $\sim$10 Myr.
Other cosmological simulations \citep{Ma17, Rosdahl18, Finlator18} show a reasonable agreement in the mean SFMS.

The scatter of the SFMS increases towards lower masses. 
At low masses, $\Ms\simeq 10^7 \Msun$, the scatter is significantly higher,  $\sigma=0.6$ dex.
This mass-dependent scatter is clearly visible in the $\sSFR-\Ms$ plane (bottom panel of \Fig{MS}).
It is mainly driven by a population of low-mass, $\Ms \le 10^8 \ \Msun$, quiescent galaxies with $\sSFR \le 10^{-0.5} \GyrI$. This population forms a tail of low sSFR-values.
Interestingly, there are no quiescent galaxies at higher masses, which implies that the number density of this hypothetical population should be lower than the limit of this sample ($\sim 3 \times 10^{-4} h^{-1} \Mpc^{-3}$).
These findings agree well with the simulations by \cite{Ma17}. They argue that  the scatter is higher at lower masses as a result of stronger burstiness in their SF histories. 
However, their results do not quantify this burstiness or provide any clue about the mechanism behind these phenomena. We will see below that the analysis of the gas content of these galaxies is crucial for the understanding of the mass-dependent scatter of the SFMS.


In order to understand the mechanisms behind the large variations in sSFR, we first divide the $\sSFR-\Ms$ plane into two-dimensional bins and compute the gas fraction (i.e. the mass in gas divided by the total mass in gas and stars)
in each bin at $z\simeq6$ (\Fig{tD}).
We see two characteristic trends. 
At a fixed $\sSFR\simeq 5 \GyrI$,
the gas fraction decreases with increasing stellar mass, from $f_G\simeq0.8$ at $\Ms\simeq 10^6 \ \Msun$ to 0.5 at $\Ms\simeq 10^9 \ \Msun$. 
Therefore, low-mass galaxies have relatively more gas available for more intense bursts with higher sSFRs.

A high gas content is needed to generate high sSFR.  Even at a fixed stellar mass,
galaxies with higher sSFR have also higher gas fractions.
The mechanisms responsible for these high sSFRs should involve the accretion of large amounts of gas. 
On the other hand, galaxies with low sSFR also show  low gas fractions.
This implies that any mechanism, such as feedback-driven outflows, able to deplete or expel significant amounts of gas will naturally lead to low sSFRs. 

The gas depletion time, $t_D=M_G / {\rm SFR}$, decreases as the gas fraction decreases at fixed sSFR, as expected (\Fig{tD}). 
Typical values are much shorter than the Hubble time, requiring the supply of new accreted gas for the maintenance of these SFR values. 
The trend at fixed stellar mass is more interesting.
The depletion time is shorter at higher sSFRs. 
As shown above, higher sSFRs also imply higher gas masses but this is not able to compensate for the rapid increase in SFR.
This trend is unexpected if the increase in sSFR at fixed mass is only driven by higher gas masses and higher gas accretion rates. 
Very dissipative processes able to increase the gas densities and decrease the gas consumption timescales are needed. 
Indeed, gas-rich mergers with a median ratio of 1:5 are found in the majority of galaxies at the top edge of the SFMS (0.5 dex above the average sSFR at all masses).
These are the best candidates to explain the bursts of SF, shown in \Fig{SSFHStrong}, although  the full characterization of the merger histories of these galaxies is beyond the scope of this paper.

The combination of these trends in gas fraction and depletion time generates the mass-dependent scatter. 
The lowest gas fractions appear in the most massive galaxies with the lowest sSFR (bottom-right corner of \Fig{MS}).
In the limit of zero gas, there is no SF and this is the reason why there are no massive galaxies with very low sSFR.
At lower masses (V$_{\rm max} \leq 100 \kms$), feedback is more efficient in ejecting gas and preventing SF \citep{DekelSilk86}.
The strong outflows generated by these violent processes are able to eject a significant fraction of  gas, decreasing the gas fraction and the sSFR and increasing the gas depletion time.

\section{Properties of SF Bursts at redshift 6}
\label{sec:Bursts}

\begin{figure}
	\includegraphics[width=\columnwidth]{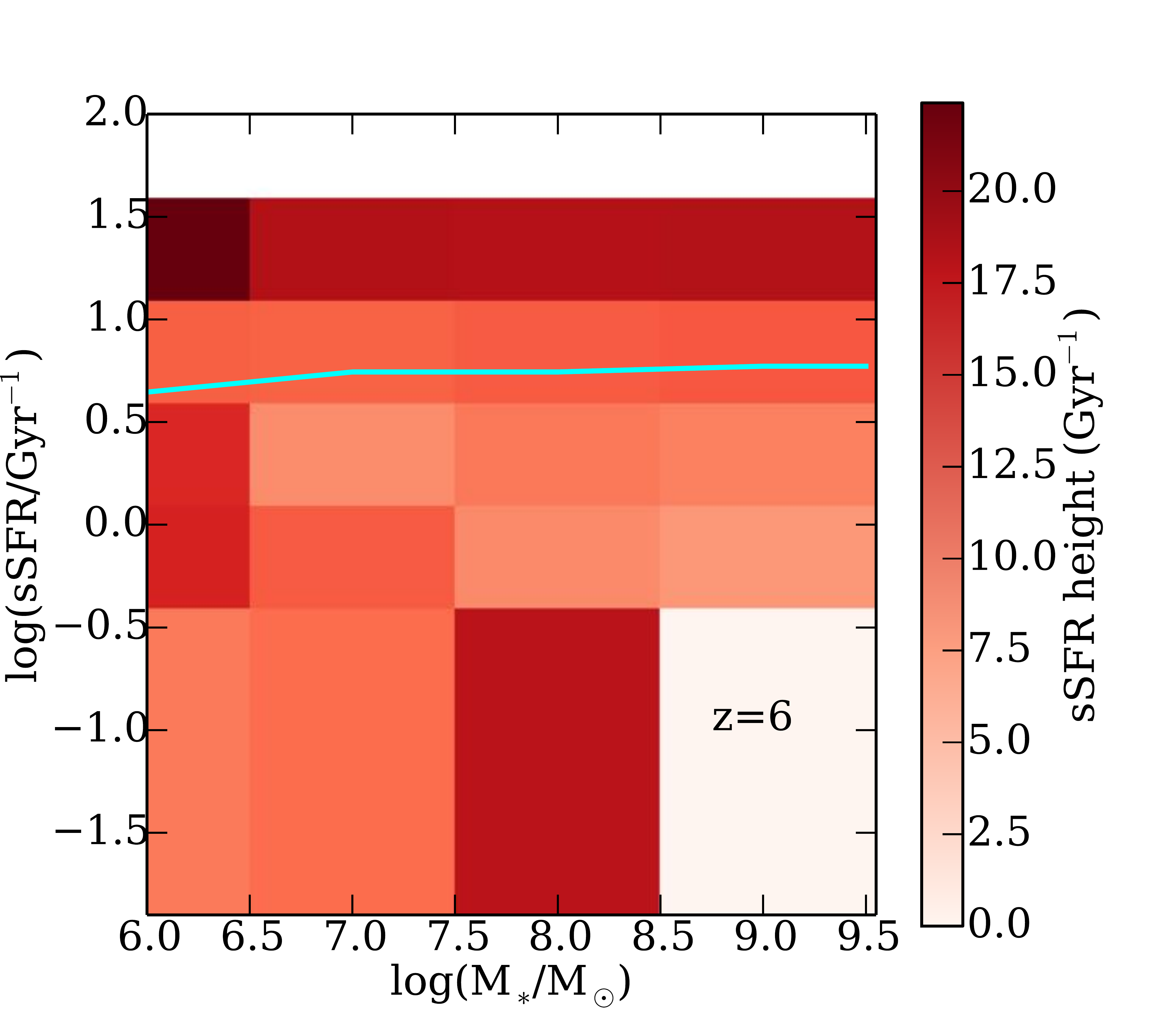}	
	 \caption{SF burst heights. 
	 The bulk of the population around the SFMS have bursts of SF with typical maxima around  $\sSFR\simeq (10 \pm 5) \GyrI$. A quarter of the population show much stronger bursts with typical heights about $\sSFR\simeq (20 \pm 5) \GyrI$. 
	 The cyan line represents the mean sSFR at different masses.}
	 \label{fig:Aheight}
\end{figure}

\begin{figure}
	\includegraphics[width=\columnwidth]{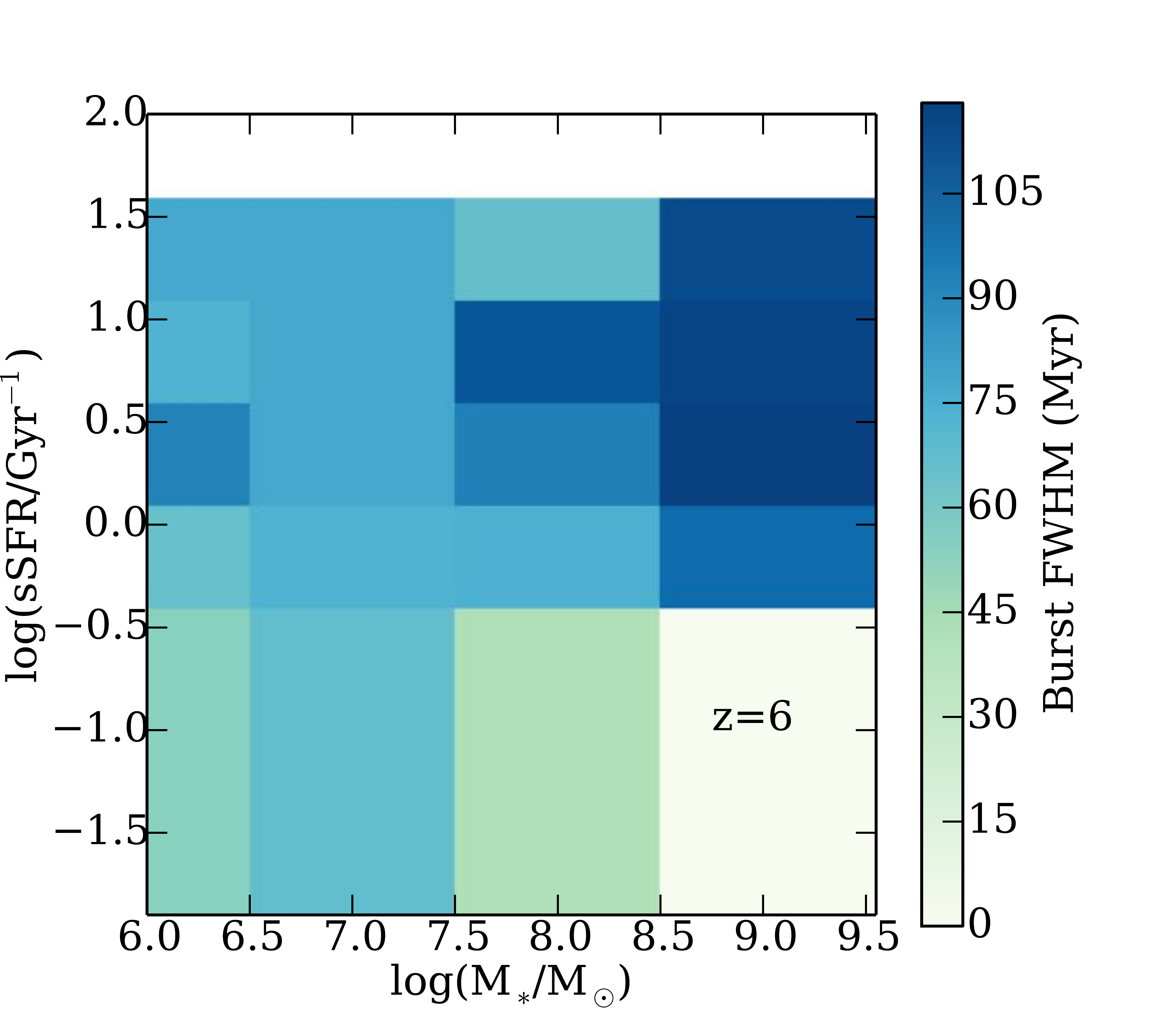}
	\includegraphics[width=\columnwidth]{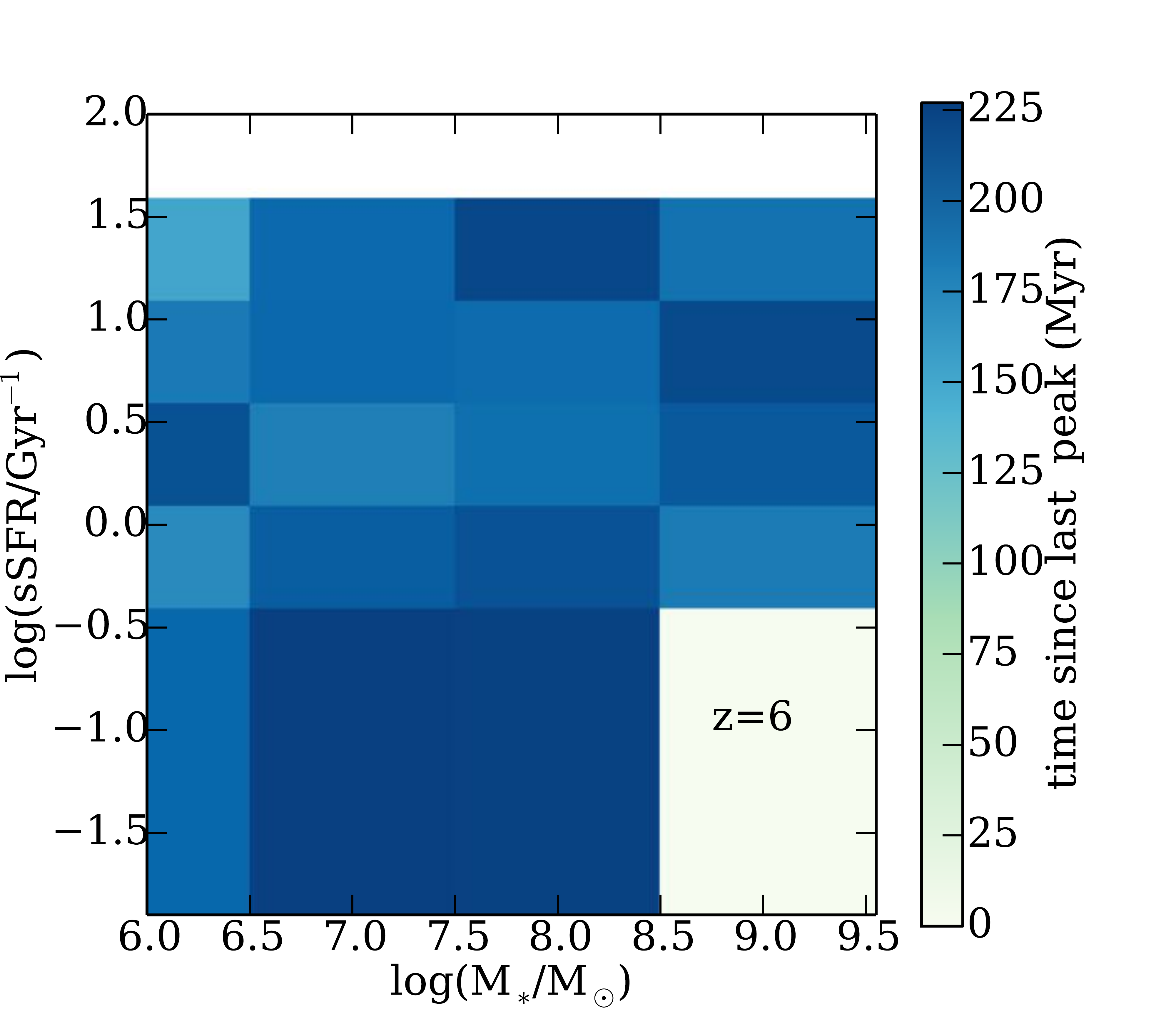}	
	\caption{FWHM of the bursts (top), and time since the last burst (bottom). There is a mild increase in both properties with stellar mass, consistent with dynamical friction and mass accretion. This is consistent with the idea  that the SF bursts are mainly driven by accretion events, modulated by feedback.}	
	 \label{fig:Btime}
\end{figure}

In this section, we focus on the properties of the SF bursts at $z\simeq6$. 
We use the snapshots of the previous section.
In each snapshot, we check if a SF burst is ongoing. This is quite common as described in Section \se{SFR}.
Then, we compute the mean properties of  SF bursts in each bin in the  $\sSFR-\Ms$ plane. 

\Fig{Aheight} shows the mean sSFR height, defined as the maximum sSFR during the burst. 
In general, this is higher than the sSFR at a given snapshot within the burst (y-axis in \Fig{Aheight}).
Only when the snapshot coincides with the peak of the burst does its sSFR equal the sSFR height.

Galaxies at the top edge of the SFMS (the highest sSFR at any mass scale) are undergoing the strongest bursts with  $\sSFR\simeq 20 \GyrI$, as expected. 
A quarter of the population at $\simeq6$ show these bursts; a few examples are shown in \Fig{SSFHStrong}. 
Interestingly, some snapshots with very low sSFR and $\Ms\simeq 10^8 \ \Msun$ also have similarly high sSFR heights.
This is a population of galaxies that undergo one of these strong bursts in their past. The burst expels significant amounts of gas, leaving behind a gas poor (25\% gas fraction), low-mass quiescent galaxy. 
The majority of galaxies along the SFMS have bursts with typical maxima of $\sSFR\simeq  (10 \pm 5) \GyrI$ independent of mass. 
This lack of mass dependence translates into a relatively tight SFMS spread over more than 3 orders of magnitude in stellar mass. 

\Fig{Btime} shows the mean FWHM as a function of mass and sSFR.
The duration of the bursts increases from FWHM$\simeq 80$ Myr at $\Ms\simeq  10^7 \Msun$ to
FWHM$=115$ Myr at $\Ms \simeq  10^9  \Msun$.
This mild mass dependence is consistent with the predictions of dynamical friction (e.g. equation 7.26 of Binney \& Tremaine 1987).  The orbital decay of two merging galaxies depends mostly on the merger mass ratio, as shown in large cosmological simulations \citep{Snyder17}.  
This suggests that SF bursts are mainly driven by gas-rich accretion events, modulated by feedback.

The increase in the scatter of the SFMS at lower masses is driven by galaxies well below this main sequence.
These galaxies have experienced bursts with heights of $\sSFR\simeq 15-20  \GyrI$, higher than the mean height of the overall population.  
At the same time, these bursts have a significantly shorter duration of FWHM $\simeq 50$ Myr at all masses. 
We thus conclude that short  and intense SF bursts are able to eject significant amounts of star-forming gas due to feedback-driven outflows. This quenches SF for some time, until new or re-accreted gas replenishes the supply of gas available for star formation.

The bottom panel of 	 \Fig{Btime} shows the mean period of time between consecutive bursts.
This period is typically $\sim$200 Myr, a significant fraction of the age of the Universe at $z=6$.
However, the distribution of values is wide. The typical dispersion within a given bin is about 100 Myr.
We do not see a strong trend with mass or sSFR.
There is only a mild mass dependence for galaxies along the SFMS.
The period between bursts increases from $\sim$180 Myr for   $\Ms\simeq  10^7  \Msun$ to $\sim$215 Myr for $\Ms\simeq  10^9  \Msun$.
This may be a consequence of the mild mass dependence of the merger rate, as shown in simulations \citep{NeisteinDekel08, Snyder17}.

\section{Evolution of SF properties}
\label{sec:evo}

\begin{figure}
	\includegraphics[width=\columnwidth]{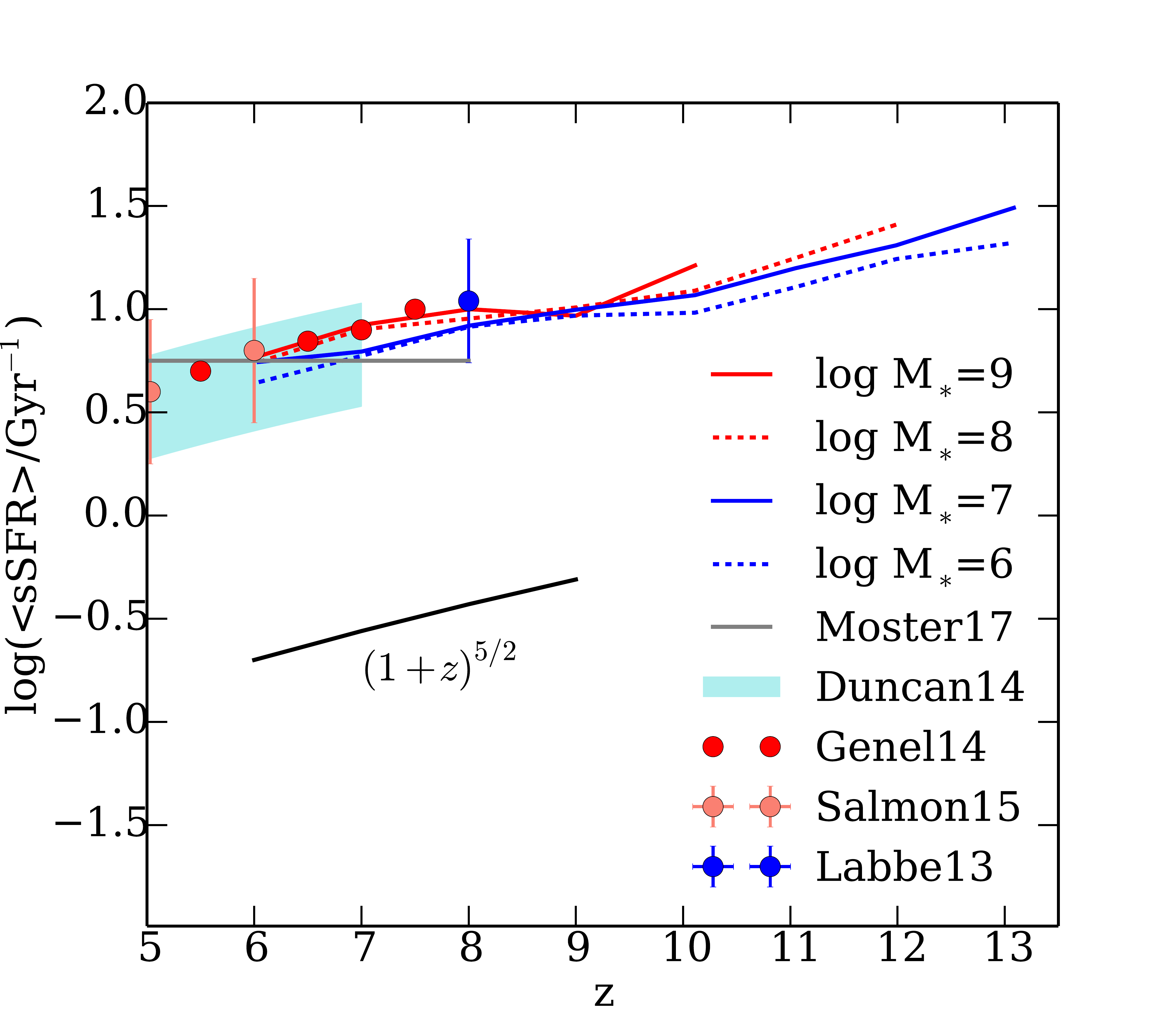}
	\caption{Evolution of the mean sSFR at different masses.
	The sSFR increases with redshift at all masses.
This is a natural prediction of the CDM paradigm of structure formation and is consistent 
with observations at $z\leq8$.}
	\label{fig:evoSSFR}
\end{figure}

In this section we focus on the evolution of the SF bursts with time.
\Fig{evoSSFR} shows the evolution of the mean sSFR at fixed stellar mass.
The sSFR increases with redshift at all masses, $\sSFR \propto (1+z)^{5/2}$, as predicted by the EPS formalism \citep{DekelMandelker14}.
In contrast, current abundance-matching models \citep{Behroozi13,Moster17} show a lack of evolution at these redshifts, 
still consistent with some observations \citep{Duncan14}, but their trend is inconsistent with current data.
They use observations with a strong overestimation of the stellar mass due to photometric contamination by emission lines \citep{Gonzalez11, Stark13}. 
Current observations at $z \leq 8$ \citep{Labbe13, Duncan14, Salmon15},
and other simulations \citep{Genel14} 
are consistent with the evolution and the overall normalization predicted by FirstLight.
At higher redshifts, simulations show even higher values with a mean of $\sSFR \simeq 30 \GyrI$ at $z\simeq13$.
This is up to an order of magnitude higher than for galaxies of similar mass at $z=5$.
The scatter around the mean sSFR is 0.3-0.6 dex, depending on the stellar mass, as discussed in \se{SFR}.



The typical sSFR height of the SF bursts also increases with redshift, 
in such a way that it is always a constant factor higher than the mean sSFR at a fixed mass and redshift: 
$\sSFR_{\rm max} / \sSFR_{\rm mean}(z)=2 \pm 0.5$ at $z=6-13$.
Appendix A shows the evolution of the SFMS from $z=15$.
The typical sSFR maximum in a SF burst at $z=9-10$ is $\sSFR_{\rm max}=20-30 \GyrI$. 
The tail of the distribution reaches even higher values of
$\sSFR_{\rm max}\simeq60 \GyrI$, for $\Ms\simeq3 \times 10^8 \Msun$ and halo masses of $\Mv \simeq 2 \times 10^{10} \Msun$ at these high redshifts.

The evolution in the sSFR could in principle be due to an evolution in the gas fraction.
However, we do not see such evolution at fixed mass. 
Low-mass galaxies, $\Ms\simeq 10^6 \Msun$, remain with $f_G\simeq0.7-0.8$ at all redshifts.
On the other hand, high-mass galaxies,   $\Ms\simeq 10^9 \Msun$, have significantly lower gas fractions, $f_G\simeq0.4-0.5$.
This lack of evolution between $z=6-13$ is also shown in the solutions of the bathtub model \citep{DekelMandelker14} if the baryonic accretion is dominated by gas rather than stars.  
It is the result of the balance between gas accretion, star formation and outflows.

The evolution of the sSFR is mainly driven by the shorter timescales at higher redshifts, due to the fact that the Universe is smaller and denser  at these epochs. All processes are more intense and violent at early times.
All timescales related with SF: gas depletion time, burst width and the time between bursts, scale with the age of the Universe.
For example, the gas depletion time can be as short as 70 Myr for galaxies with $\Ms\simeq 10^9 \Msun$ at $z=10$, when the Universe is only 500 Myr old.
At that time, galaxies with $\Ms\simeq10^7 \Msun$ have a burst width of only 40 Myr and the time between bursts is 100 Myr, still a fifth of the age of the Universe at that redshift.

\section{Conclusions and Discussion}
\label{sec:summary}

We have used the FirstLight database (paper I) to study the SF histories of $\sim$300 galaxies with a stellar mass between $\Ms=10^6$ and $3 \times 10^9 \Msun$ during cosmic dawn ($z=5-15$). 
The main results can be summarized as follows:
\begin{itemize}
\item
The evolution of the SFR in each galaxy is complex and diverse, characterized by bursts of SF. 
Overall, first galaxies spend about 70\% of their time undergoing SF bursts at $z>5$.
\item
This diversity sets the mean and scatter of the SFMS at $z\simeq5-13$.
\item
A mass-dependent scatter is driven by a population of low-mass, $\Ms \leq 10^8 \Msun$, quiescent galaxies.
\item
High gas fractions and short gas depletion times are common during the SF bursts.
\item
The typical bursts at $z\simeq6$ have a sSFR maximum of  $5-15 \GyrI$ with a FWHM$\sim$100 Myr, one tenth of the age of the  Universe.
\item
A quarter of the bursts populate a tail with very high sSFR maxima of $20-30 \GyrI$ and significantly shorter time-scales of FWHM $\sim 40-80$ Myr at all masses.
\item
The mean period of time between consecutive bursts is $\sim$200 Myr with a small mass dependence at $z\simeq6$. 
\item
The mean sSFR increases with redshift approximately as $\sSFR \propto (1+z)^{5/2}$ at all masses, as predicted by $\Lambda$CDM models. This is consistent with existing observations at $z\leq8$.
\item
The typical sSFR height of a SF burst also increases with redshift, but it is always a factor $2\pm0.5$ higher than the mean sSFR at that redshift.
This implies typical sSFR maxima of $\sSFR_{\rm max}=20-30 \GyrI$ at $z=9-10$. The tail of the distribution reaches $\sSFR_{\rm max}\simeq 60 \GyrI$ at these high redshifts.
\item
This evolution is driven by shorter time-scales at higher redshifts, proportional to the age of the Universe.
There is no evolution in the gas fraction at fixed stellar mass.
\end{itemize}

One of the caveats in the present analysis is the limitation to SF bursts longer than 30 Myr.
Bursts shorter than this limit are difficult to distinguish from small fluctuations in the SF history. 
This implies that the sSFR maximum of these short bursts is underestimated.
In practice,  the number of bursts with a width close to this limit is very small.
The fact that we do not see significant and short bursts in our simulations does not imply that they do not exist in the early Universe. 
In fact, there are samples of Ly$\alpha$ emitters younger than $\sim$30 Myr, with masses between 
$\sim 10^8 - 10^9 \Msun$ and $ \sSFR  \simeq  60 \GyrI$ at $z\simeq6$
  \citep{Jiang16}. We may need a larger sample in bigger cosmological volumes to see such rare events. 

The SF histories are publicly available at the FirstLight website\footnote{\url{http://www.ita.uni-heidelberg.de/~ceverino/FirstLight}}, along with other galaxy properties such as the virial radius, the halo mass, or the galaxy positions within the simulation box. 
This database is complemented with the spectral energy distributions for all snapshots every 10 Myr, as described in a companion paper (Ceverino et al. in prep.)
This data release will be the base for future mock JWST surveys that will help with the planning and the interpretation of future observations in the coming decade with JWST and large ground-based 30 m telescopes.

\section*{Acknowledgements}

We thank the anonymous referee for his/her comments which improve the quality of this paper.
We thank Annalisa Pillepich and Linhua Jiang for fruitful discussions.
This work has been funded by  the ERC Advanced Grant, STARLIGHT: Formation of the First Stars (project number 339177).
RSK and SCOG also acknowledge support from the DFG via SFB 881 `The Milky Way System' (sub-projects B1, B2 and B8) and SPP 1573 `Physics of the Interstellar Medium' (grant number GL 668/2-1).
The authors gratefully acknowledge the Gauss Center for Supercomputing for funding this project by providing computing time on the GCS Supercomputer SuperMUC at Leibniz Supercomputing Centre (Project ID: pr92za).
The authors acknowledge support by the state of Baden-Württemberg through bwHPC.




\bibliographystyle{mnras}
\bibliography{SFR5} 


\appendix

\section{The Star-Forming Main Sequence at different redshifts}

\begin{figure}
	\includegraphics[width=\columnwidth]{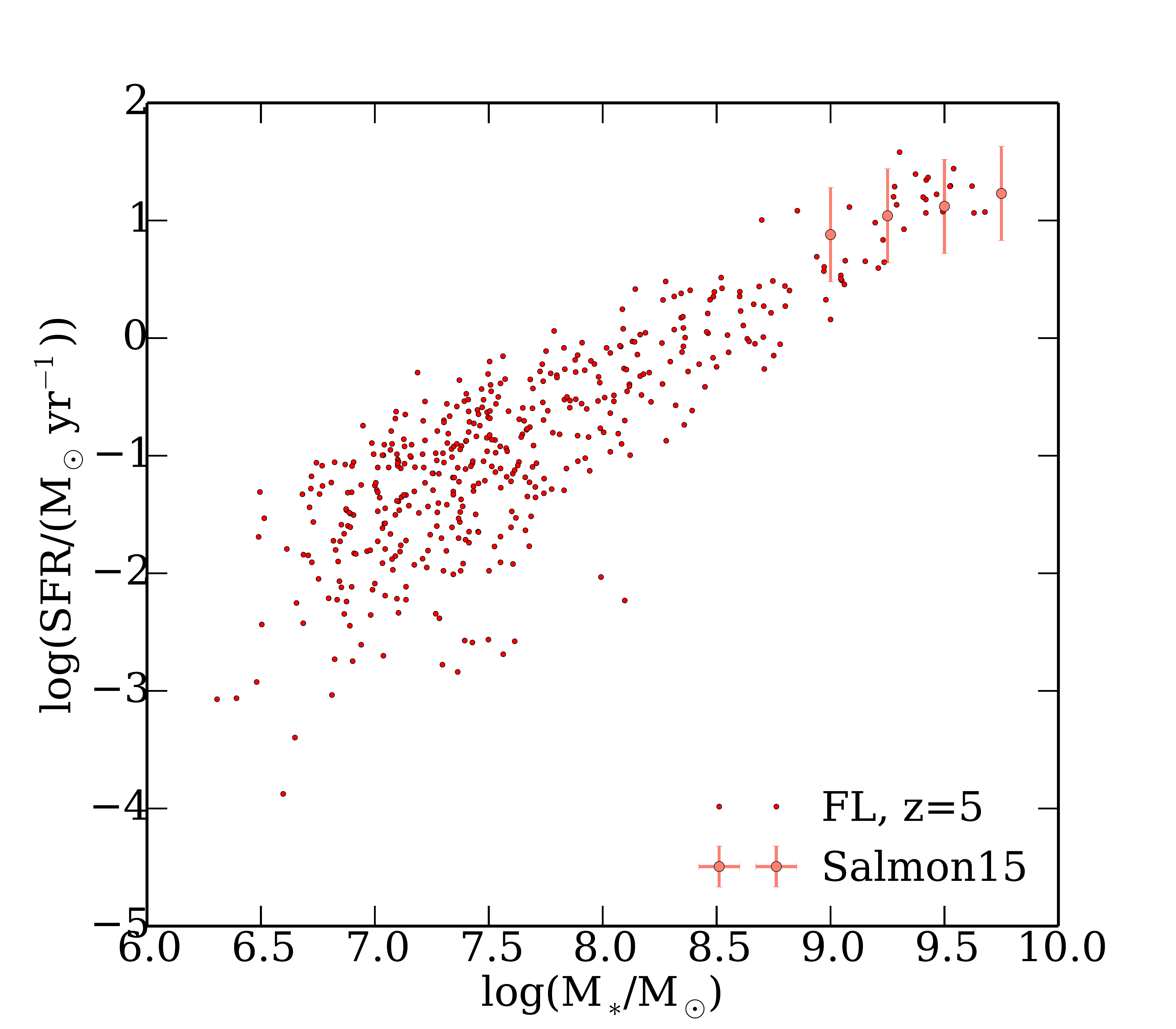}

	 \caption{SFR versus stellar mass ($\Ms$) at z=5.}
	  \label{fig:MSz5}
\end{figure}

\begin{figure}
	\includegraphics[width=\columnwidth]{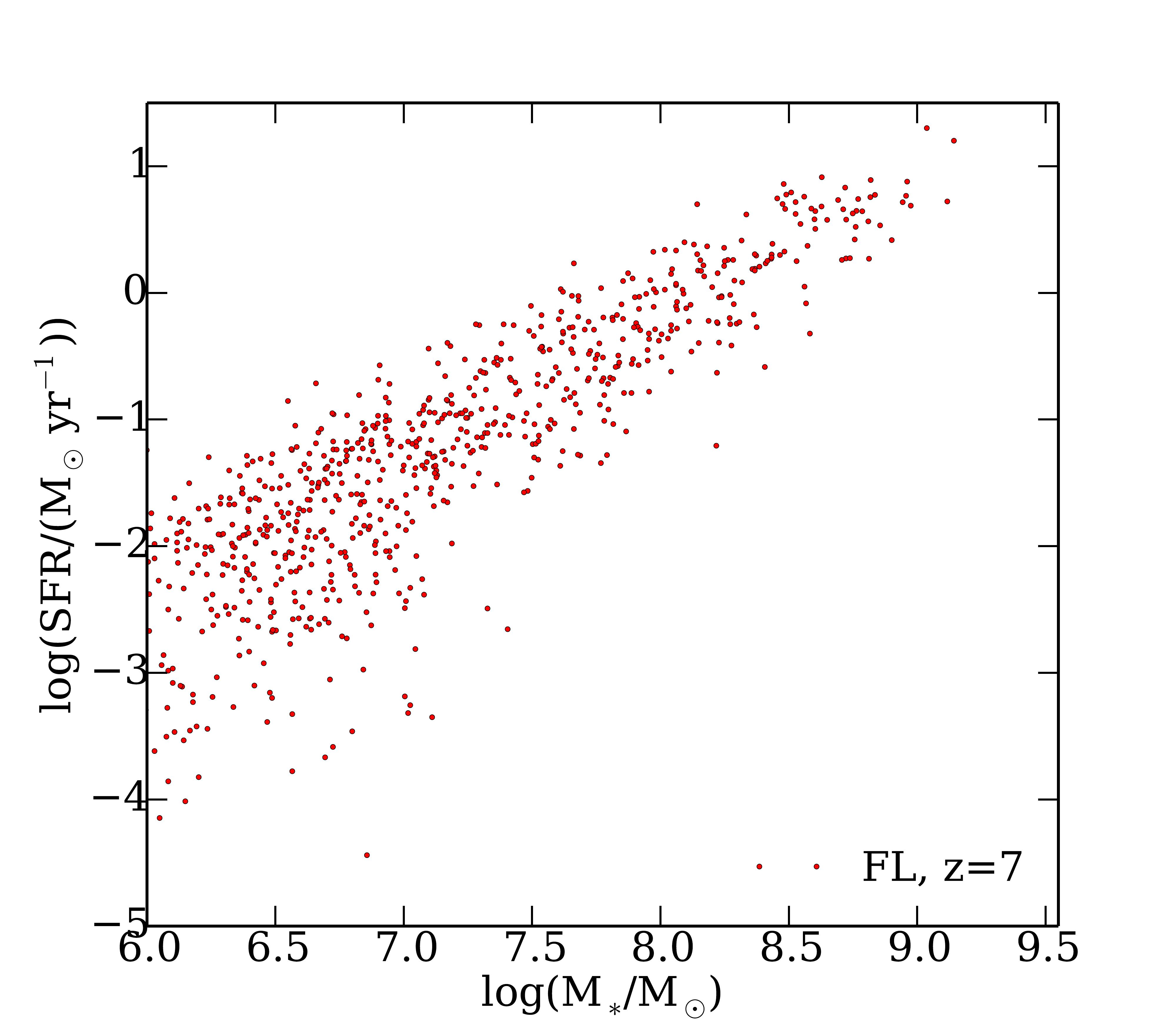}
	\includegraphics[width=\columnwidth]{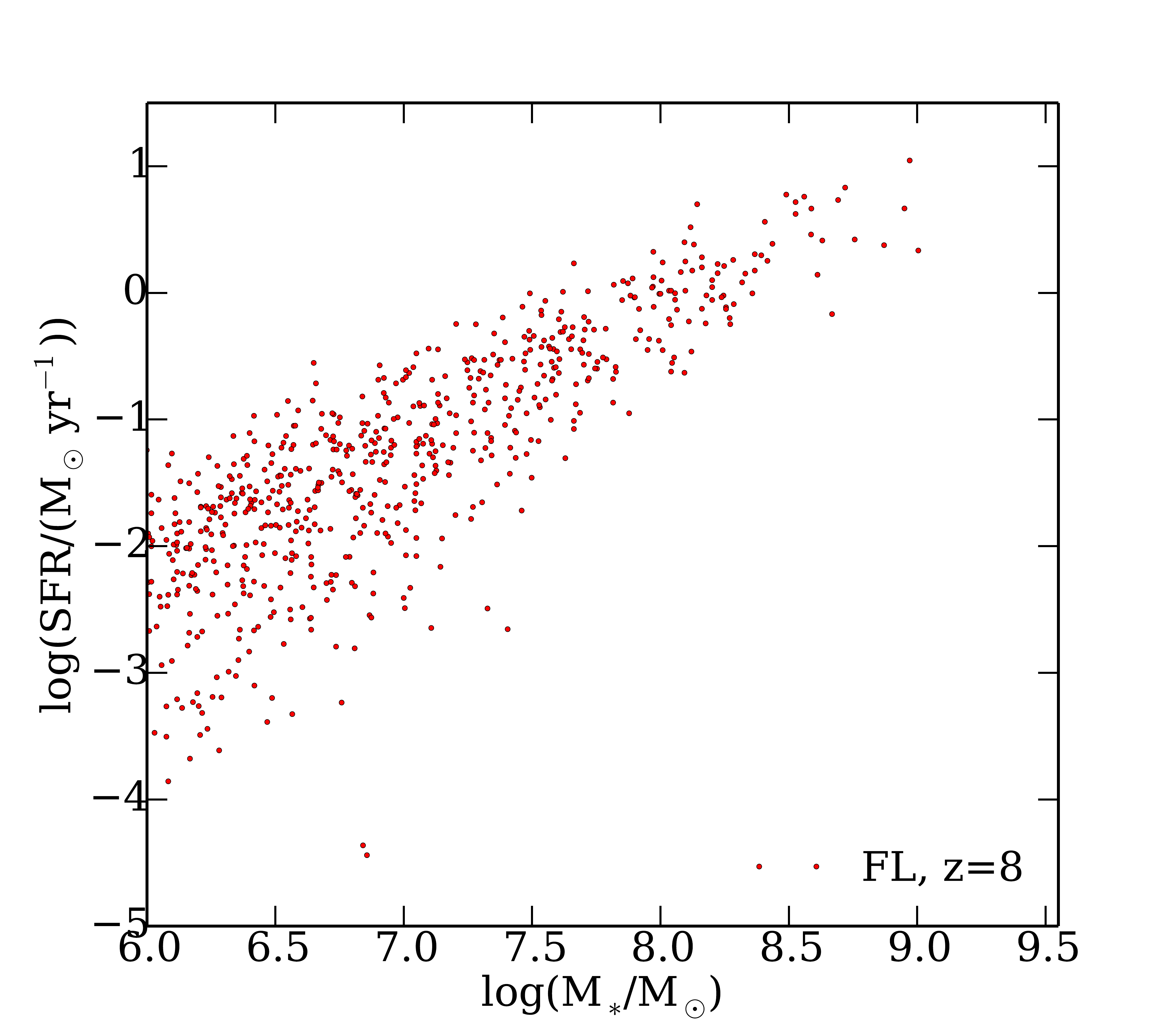}	
	\includegraphics[width=\columnwidth]{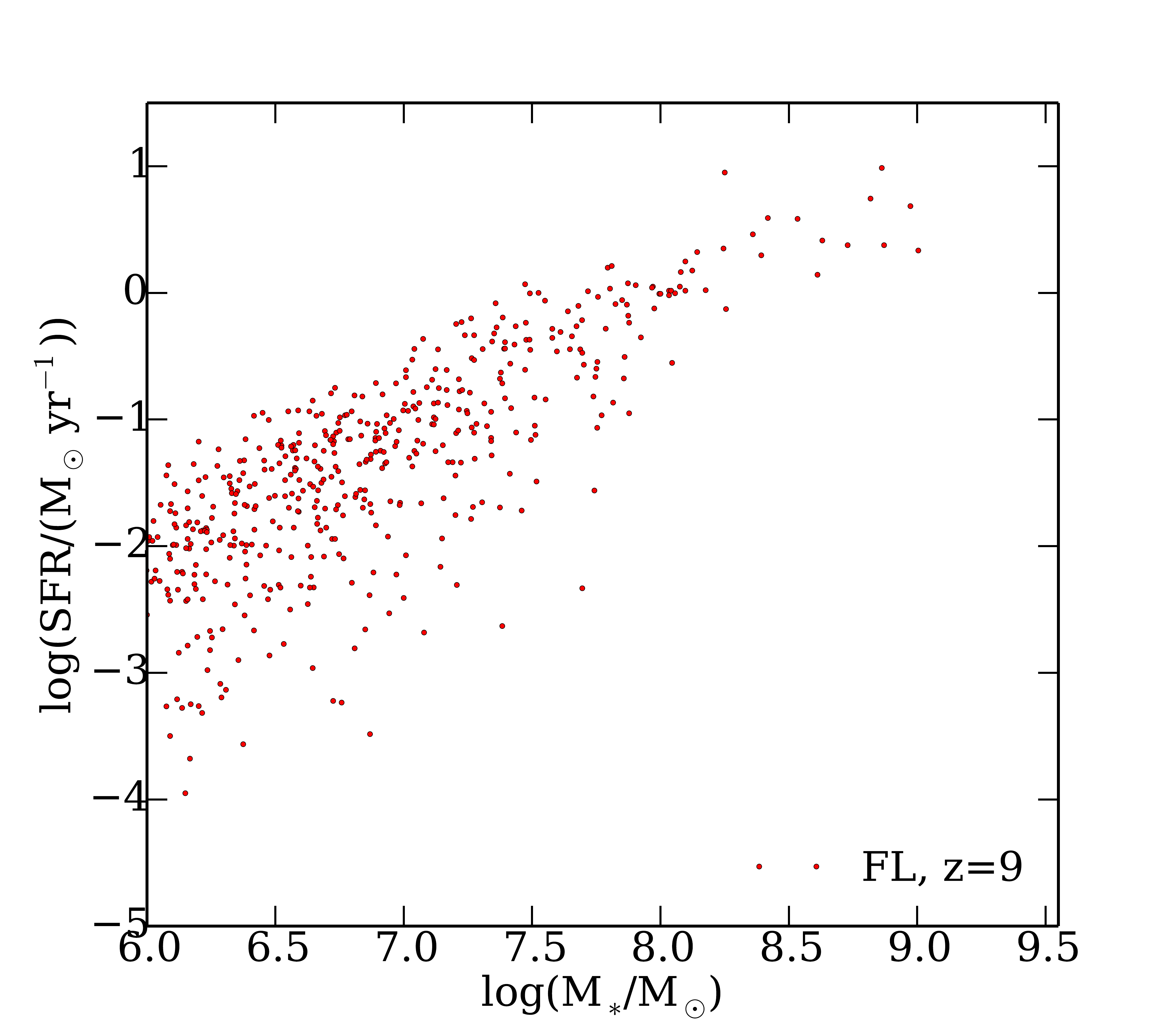}	

	 \caption{SFR versus stellar mass ($\Ms$) at z=7,8,9.}
	  \label{fig:MSz79}
\end{figure}

\begin{figure}
	\includegraphics[width=\columnwidth]{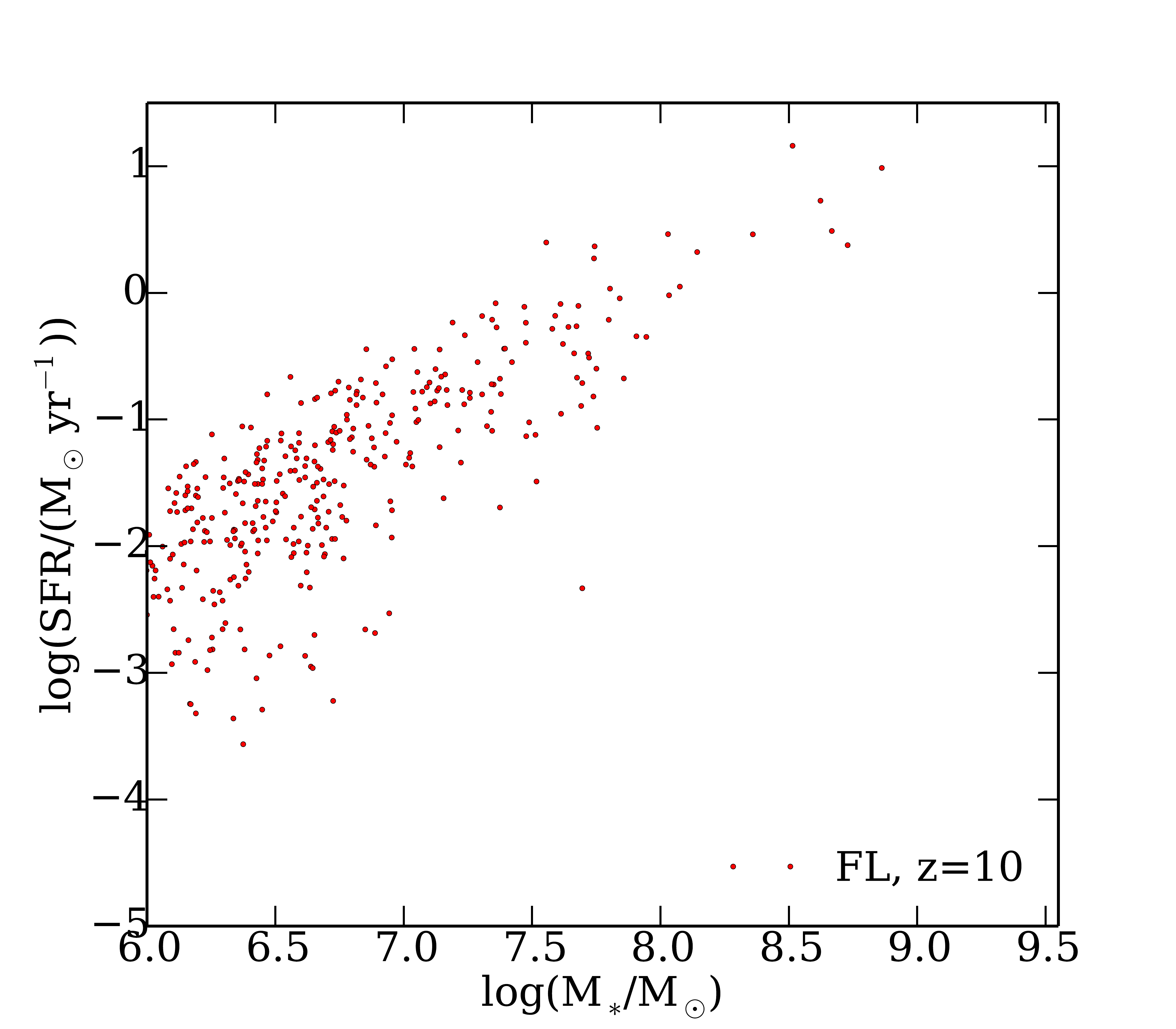}
	\includegraphics[width=\columnwidth]{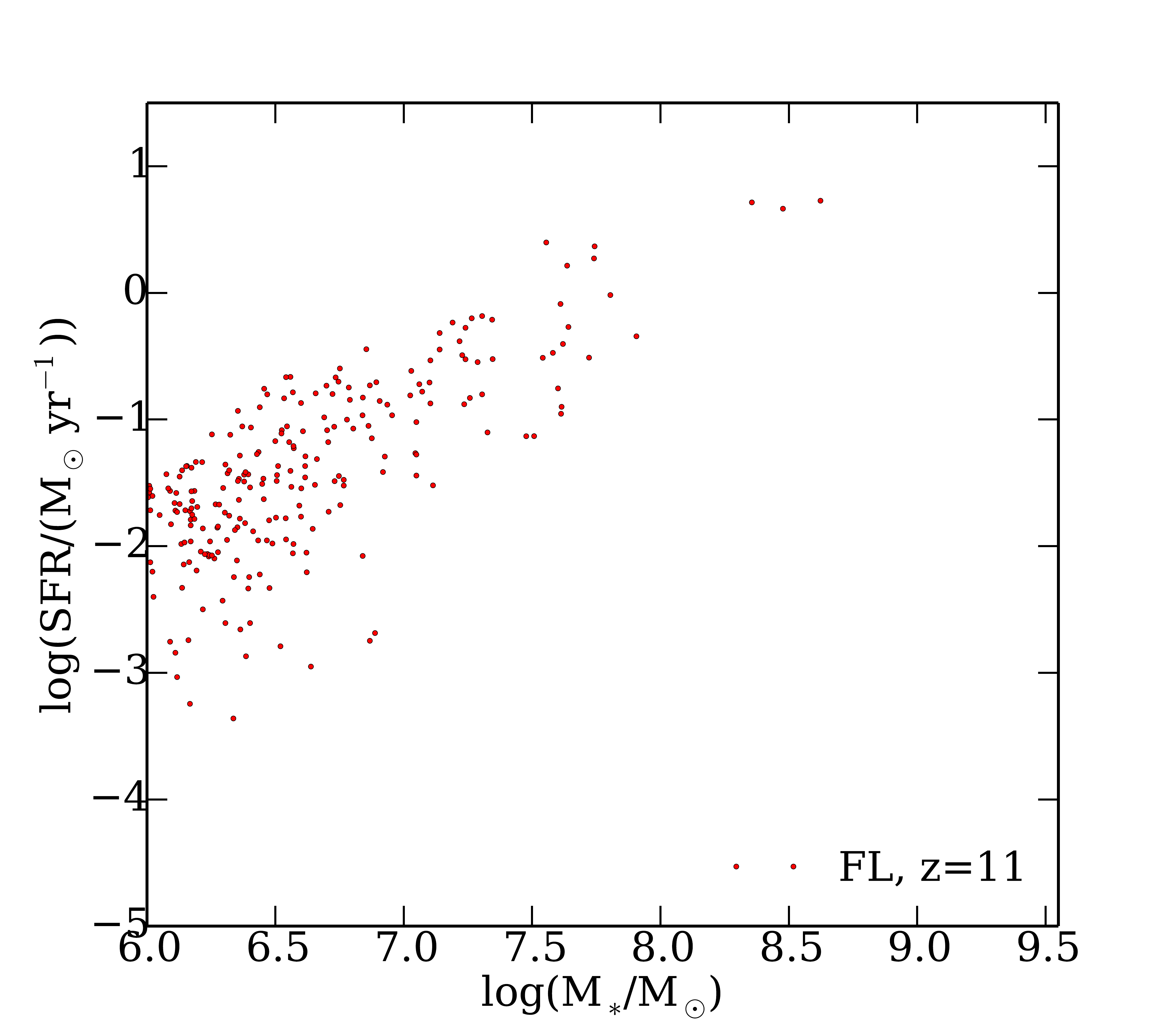}	
	\includegraphics[width=\columnwidth]{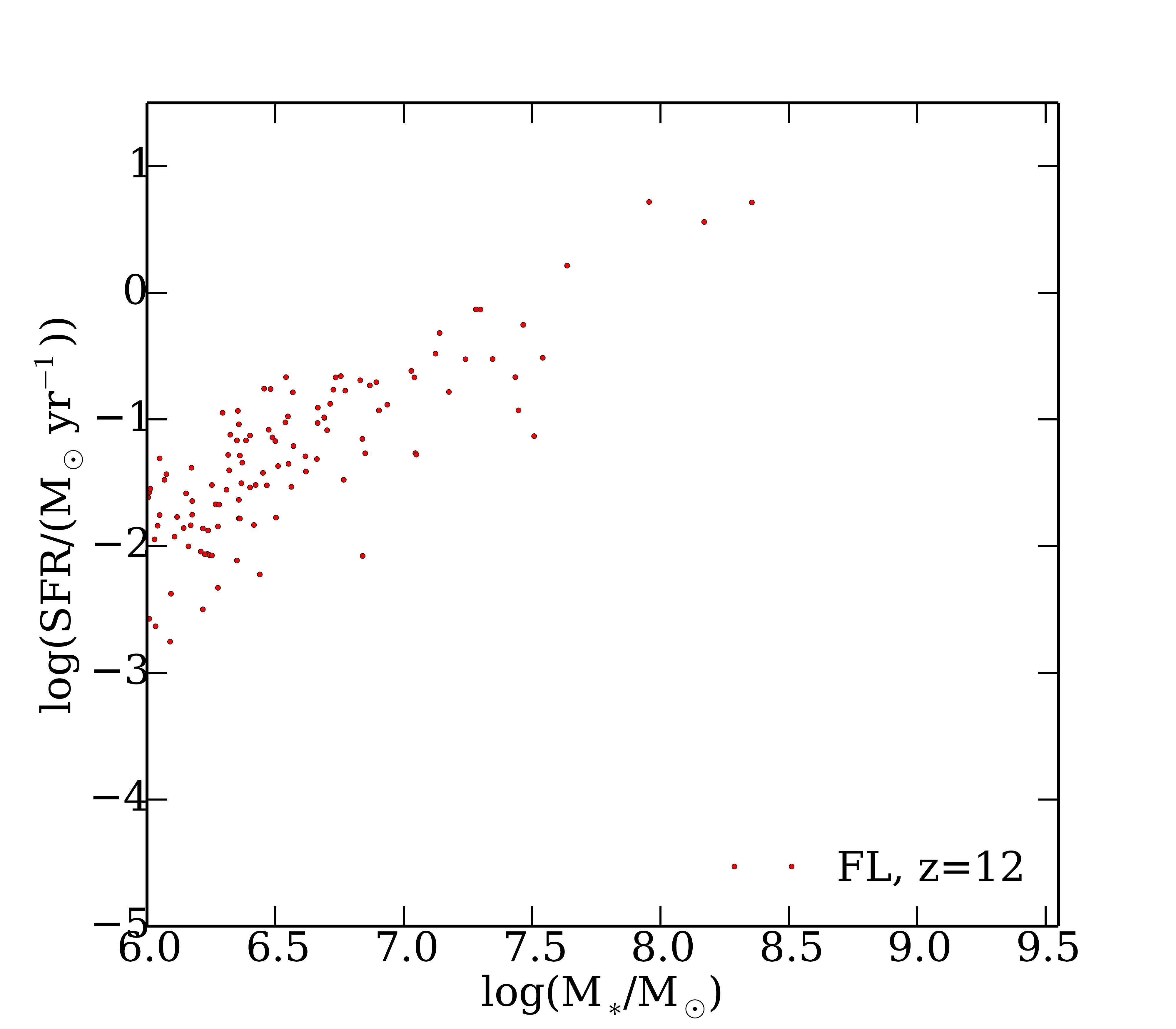}	

	 \caption{SFR versus stellar mass ($\Ms$) at z=10,11,12.}
	  \label{fig:MSz1012}
\end{figure}

\begin{figure}
	\includegraphics[width=\columnwidth]{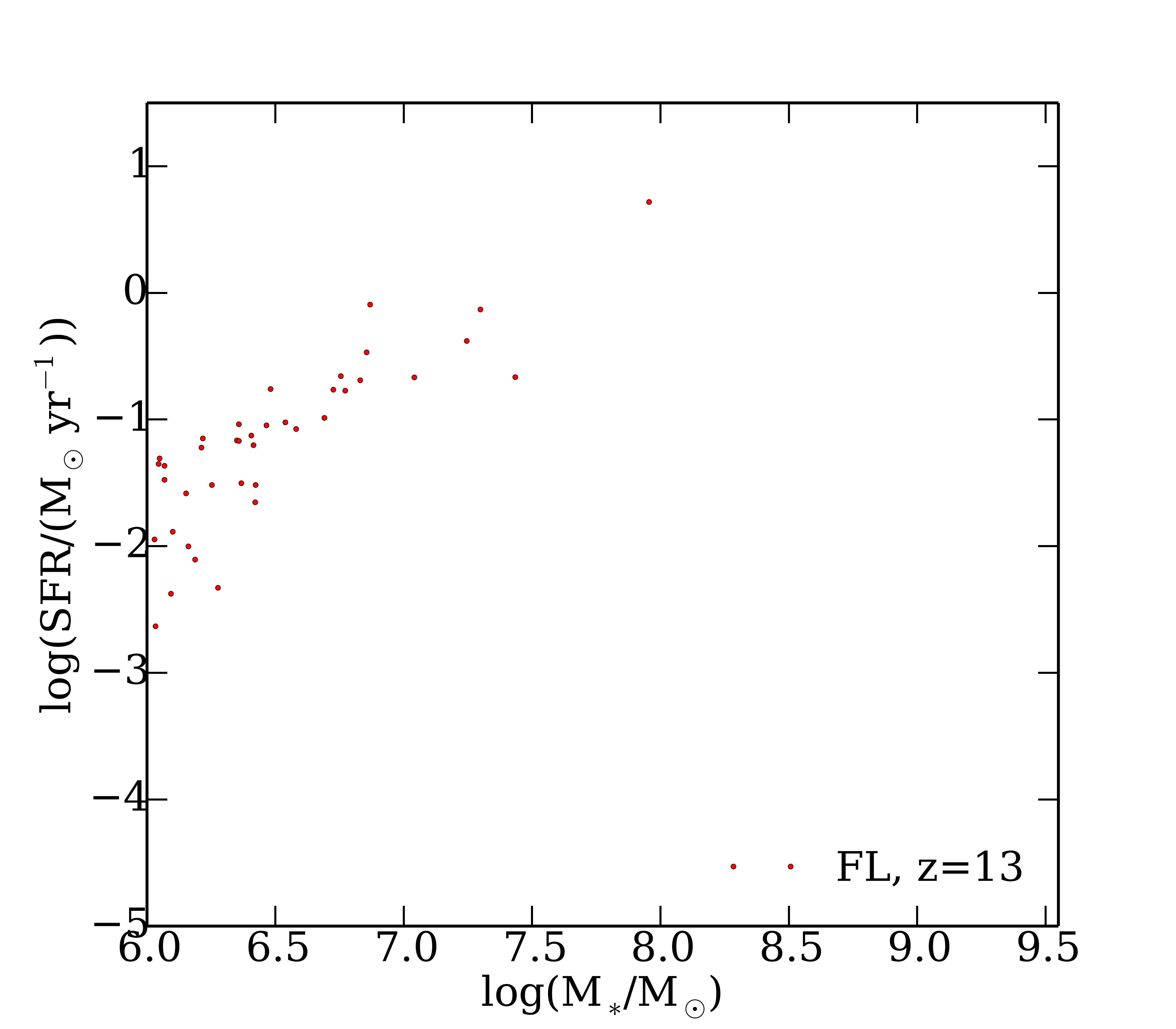}
	\includegraphics[width=\columnwidth]{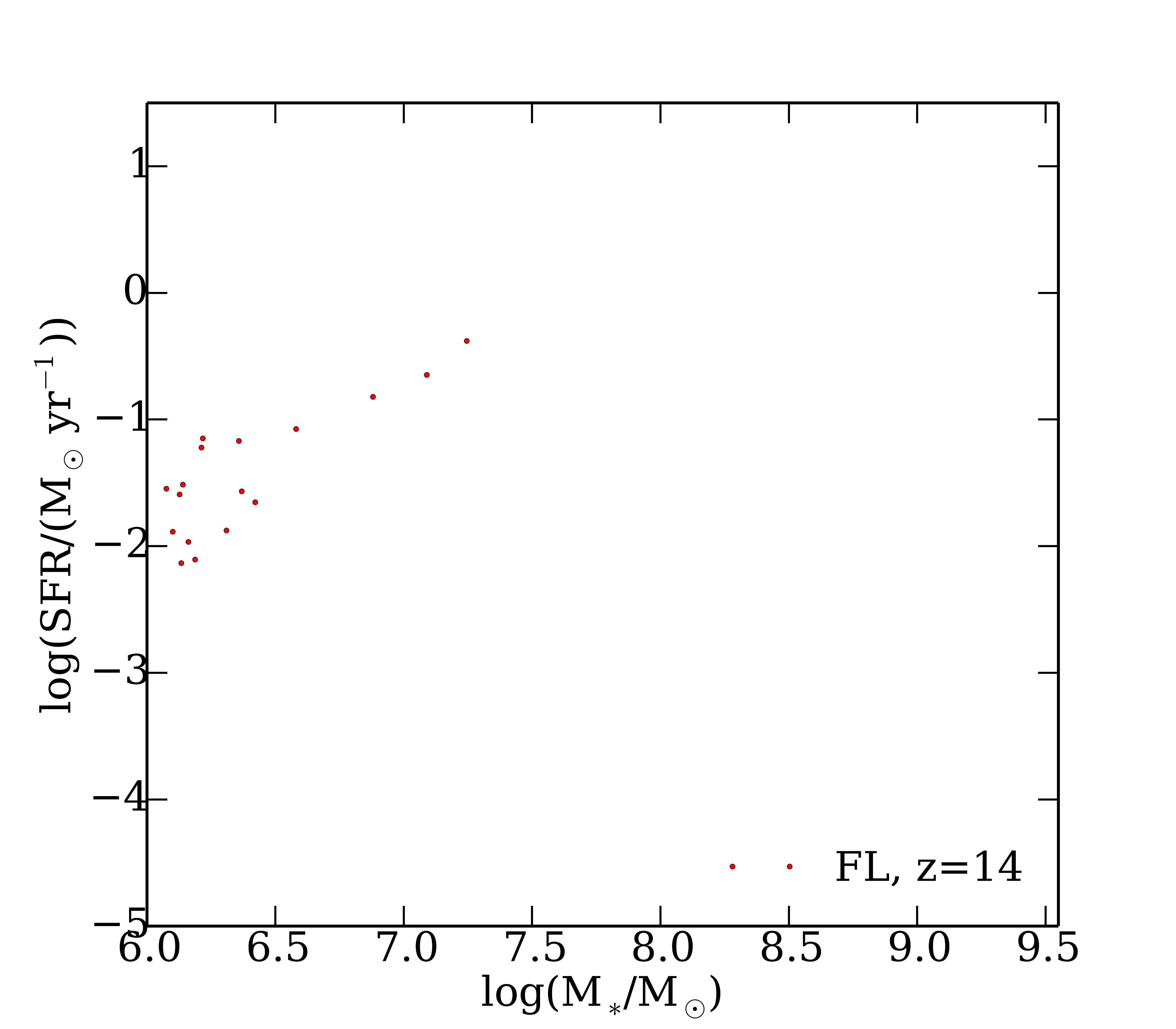}	
	\includegraphics[width=\columnwidth]{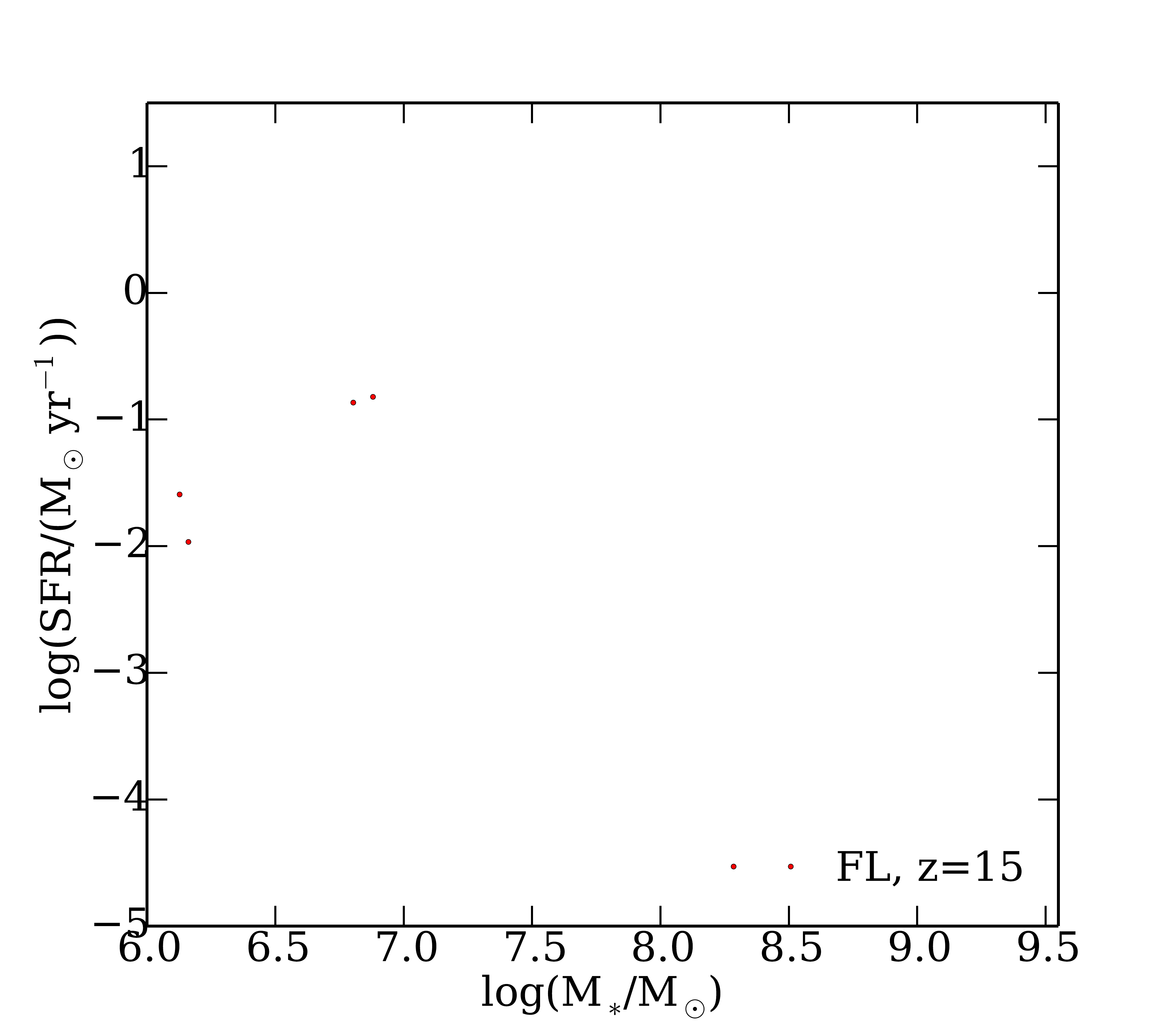}	

	 \caption{SFR versus stellar mass ($\Ms$) at z=13,14,15.}
	  \label{fig:MSz1315}
\end{figure}

\Figs{MSz5}, \ref{fig:MSz79}, \ref{fig:MSz1012}, and \ref{fig:MSz1315} show the SFMS at different redshifts.
Close to the last available snapshot, $z\simeq5$, the simulations are consistent with current observations \citep{Salmon15}.
As the redshift increases, the number of galaxies above a virial mass of $10^9 \Msun$ decreases.
This makes it difficult to estimate mean properties at redshifts higher than 13, when only a few galaxies with stellar masses higher than $10^7 \Msun$ are present within our simulated cosmological volumes. Larger cosmological volumes will be needed for more complete studies at these very high redshifts.

\bsp	
\label{lastpage}
\end{document}
